\documentclass[twocolumn]{aastex62}

\usepackage{graphicx}
\usepackage{epstopdf}
\usepackage[intlimits,sumlimits]{amsmath}
\usepackage{epsfig} 
\usepackage{amssymb}
\usepackage{mathrsfs}  
\usepackage{gensymb}

\usepackage{natbib}

\usepackage{amsmath}

\definecolor{dark-red}{rgb}{0.4,0.15,0.15}
\definecolor{dark-blue}{rgb}{0.15,0.15,0.4}
\definecolor{medium-blue}{rgb}{0,0,0.5}
\hypersetup{
    colorlinks, linkcolor={dark-red},
    citecolor={dark-blue}, urlcolor={medium-blue}
}

\newcommand{\beqa}{\begin{eqnarray}} 
\newcommand{\eeqa}{\end{eqnarray}}

\newcommand{\bsub}{\begin{subequations}}
\newcommand{\esub}{\end{subequations}}
\newcommand{\beal}{\begin{align}}
\newcommand{\ealn}{\end{align}}
\newcommand{\Nif}{$\rm ^{56}Ni$}

\newcommand{\msun}{M$_{\sun}$}

\newcommand{\ConfirmedRate}{7}

\newcommand{\DetectionRate}{14}
\newcommand{\MCSMsolmed}{0.013}
\newcommand{\MaxPhase}{1845}
\newcommand{\MaxRedshift}{0.070}

\newcommand{\MedianMagLim}{25.9}
\newcommand{\MedianPhase}{952}
\newcommand{\MedianRedshift}{0.027}
\newcommand{\MinPhase}{270}
\newcommand{\MinRedshift}{0.005}
\newcommand{\Nambiguous}{7}

\newcommand{\Ncomparison}{1}

\newcommand{\Nconfirmed}{6}

\newcommand{\Ndetected}{13}

\newcommand{\Nexcluded}{8}
\newcommand{\Nnondet}{78}

\newcommand{\Ntotal}{91}

\newcommand{\NtotalPaper}{92}
\newcommand{\NtotalSLSN}{21}
\newcommand{\dtEjectYrFiveHundred}{4}
\newcommand{\dtEjectYrHundred}{18}
\newcommand{\dtEjectYrThreeHundred}{6}

\newcommand{\fCSMerrminus}{0.09}
\newcommand{\fCSMerrplus}{0.17}
\newcommand{\fCSMmed}{0.23}

\newcommand{\fthickmed}{0.07}
\newcommand{\logMCSMerrminus}{0.48}
\newcommand{\logMCSMerrplus}{0.48}
\newcommand{\logMCSMmed}{-1.87}
\newcommand{\logfthickerrminus}{0.52}
\newcommand{\logfthickerrplus}{0.51}
\newcommand{\logfthickmed}{-1.18}

\received{\today}
\revised{}
\accepted{}
\submitjournal{ApJ}

\shorttitle{HST UV SE SNe}
\shortauthors{Fremling et al.}

\begin{document}



\title{Late-Time HST UV Detections Reveal Eruptive Mass Loss and Circumstellar Interaction in a Quarter of Stripped-Envelope Supernovae}

\correspondingauthor{C.~Fremling}
\email{fremling@caltech.edu}

\author[0000-0002-4223-103X]{C.~Fremling}
\affiliation{Caltech Optical Observatories, California Institute of Technology, Pasadena, CA 91125, USA}
\affiliation{Division of Physics, Mathematics and Astronomy, California Institute of Technology, Pasadena, CA 91125, USA}

\author[0000-0003-1858-561X]{S.~Covarrubias}
\affiliation{Division of Physics, Mathematics and Astronomy, California Institute of Technology, Pasadena, CA 91125, USA}
\author[0000-0003-1546-6615]{J.~Sollerman}
\affiliation{Department of Astronomy, Stockholm University, AlbaNova, 106 91 Stockholm, Sweden}

\author[0000-0002-8989-0542]{K.~De}
\affiliation{Department of Astronomy, Columbia University, New York, NY 10027, USA}
\affiliation{Columbia Astrophysics Laboratory, Columbia University, New York, NY 10027, USA}

\author[0000-0001-9152-6224]{T.~X.~Chen}
\affiliation{IPAC, California Institute of Technology, 1200 E. California Blvd, Pasadena, CA 91125, USA}

\author[0000-0002-1066-6098]{T.-W.~Chen}
\affiliation{Graduate Institute of Astronomy, National Central University, 300 Jhongda Road, Jhongli 32001, Taiwan}

\author[0000-0002-5884-7867]{R.~Dekany}
\affiliation{Caltech Optical Observatories, California Institute of Technology, Pasadena, CA 91125, USA}

\author[0000-0001-8532-3594]{C.~Fransson}
\affiliation{Department of Astronomy, Stockholm University, AlbaNova, 106 91 Stockholm, Sweden}

\author[0000-0002-3653-5598]{A.~Gal-Yam}
\affiliation{Department of Particle Physics and Astrophysics, Weizmann Institute of Science, Rehovot 76100, Israel}

\author[0000-0001-5668-3507]{S.~L.~Groom}
\affiliation{IPAC, California Institute of Technology, 1200 E. California Blvd, Pasadena, CA 91125, USA}

\author[0000-0002-3934-2644]{W.~V.~Jacobson-Gal\'{a}n}
\altaffiliation{NASA Hubble Fellow}
\affiliation{Cahill Center for Astrophysics, California Institute of Technology, MC 249-17, 1216 E California Boulevard, Pasadena, CA, 91125, USA}

\author[0000-0002-5619-4938]{M.~M.~Kasliwal}
\affiliation{Division of Physics, Mathematics and Astronomy, California Institute of Technology, Pasadena, CA 91125, USA}

\author[0000-0001-9454-4639]{R.~Lunnan}
\affiliation{Department of Astronomy, Stockholm University, AlbaNova, 106 91 Stockholm, Sweden}

\author[0000-0002-6786-8774]{E.~O.~Ofek}
\affiliation{Department of Particle Physics and Astrophysics, Weizmann Institute of Science, Rehovot 76100, Israel}

\author[0000-0001-8472-1996]{D.~A.~Perley}
\affiliation{Astrophysics Research Institute, Liverpool John Moores University, Liverpool L3 5RF, UK}

\author[0000-0003-1227-3738]{J.~N.~Purdum}
\affiliation{Caltech Optical Observatories, California Institute of Technology, Pasadena, CA 91125, USA}

\author[0000-0001-6797-1889]{S.~Schulze}
\affiliation{Center for Interdisciplinary Exploration and Research in Astrophysics and Department of Physics and Astronomy, Northwestern University, Evanston, IL 60208, USA}

\author[0000-0003-4531-1745]{Y.~Sharma}
\affiliation{Division of Physics, Mathematics and Astronomy, California Institute of Technology, Pasadena, CA 91125, USA}

\author{N.~Sravan}
\affiliation{Department of Physics, Drexel University, Philadelphia, PA 19104, USA}

\author[0009-0002-1319-3975]{A.~Wei}
\affiliation{Department of Astronomy, University of California, Berkeley, CA 94720, USA}

\author{Lin~Yan}
\affiliation{Caltech Optical Observatories, California Institute of Technology, Pasadena, CA 91125, USA}

\author[0000-0001-6747-8509]{Y.~Yao}
\affiliation{Miller Institute for Basic Research in Science, 206B Stanley Hall, Berkeley, CA 94720, USA}
\affiliation{Department of Astronomy, University of California, Berkeley, CA 94720-3411, USA}

\begin{abstract}
We present HST WFC3/UVIS F275W near-UV imaging of \Ntotal\ stripped-envelope supernovae (SE~SNe; Types IIb, Ib, Ic) from Snapshot program SNAP-16657, observed at phases of \MinPhase--\MaxPhase\ days (median \MedianPhase\ days) after first optical detection. We detect UV counterparts in \Ndetected\ SE~SNe, of which \Nconfirmed\ are classified as secure and \Nambiguous\ as ambiguous after comparison to nearby H\,\textsc{ii} regions, interpreting the secure sources as signatures of interaction with circumstellar material (CSM). Independent WISE W1/W2 light curves show $>300$~day mid-IR excesses in two of the secure UV sources, corroborating the interaction interpretation, and reveal two additional IR-only candidates without UV counterparts, indicating dust-obscured interaction episodes missed by the UV survey. A forward-modeling MCMC analysis using a physics-based CSM interaction model with three free parameters, the interaction fraction $f_\mathrm{CSM}$, shell mass $M_\mathrm{CSM}$, and thickness fraction $f_\mathrm{thick}$, yields $f_\mathrm{CSM} = \fCSMmed^{+\fCSMerrplus}_{-\fCSMerrminus}$, $M_\mathrm{CSM} \approx \MCSMsolmed~M_\odot$, and $f_\mathrm{thick} \approx \fthickmed$. The inferred thin-shell geometry implies an ejection duration of $\sim$\dtEjectYrThreeHundred~yr for an outflow velocity of $300$~km~s$^{-1}$, two to three orders of magnitude shorter than the thermal timescale of stable Roche-lobe overflow. This result disfavors steady binary mass transfer as the origin of the detected CSM and instead points to eruptive pre-supernova mass ejection in the final years before core collapse, either from wave-driven outbursts or from mass transfer triggered by late-stage progenitor re-expansion.
\end{abstract}



\section{Introduction}
Stripped-envelope (SE) supernovae (SNe) arise from massive stars that have lost all or most of their hydrogen envelopes before core collapse \citep{Filippenko1997,GalYam2017}. They are classified by the remaining envelope composition at the time of explosion: Type~IIb (some H), Ib (no H, some He), and Ic (no H, no He; see \citealp{Smartt2009} for a review).

Two mechanisms can strip the hydrogen envelope: strong stellar winds from very massive progenitors ($>30$~\msun; \citealp{Groh:2013ab}), or binary interactions that allow lower-mass progenitors ($<20$~\msun; \citealp{Yoon2015}) to produce SE~SNe. Observations increasingly favor the binary channel. Detailed studies of SN~1993J \citep{Schmidt1993,Nomoto1993} and SN~2011dh \citep{Arcavi2011,Ergon2014,Ergon2015}, together with larger samples \citep{Drout2011,Cano:2013aa,Taddia2015AandA574A60T,Lyman2016,Taddia2018AandA609A136T,Prentice2019,Barbarino2021}, yield ejecta masses of only a few \msun, which is too low for wind-driven stripping \citep{Groh:2013ab}. The absence of significant differences in progenitor mass or metallicity among the main SE~SN subtypes \citep{Sanders2012,Kuncarayakti2018} further supports a binary origin.

If binary orbital parameters rather than progenitor mass or metallicity\footnote{Higher metallicity increases mass loss from winds, affecting both single- and binary-star evolution.} drive the SE~SN subtype sequence, a key variable is the evolutionary stage at which mass transfer begins (see \citealp{Smith2014} for a review). Case~A \citep{Morton1960} and Case~B mass transfer strip the envelope long before core collapse, leaving little material near the star at the time of explosion. Case~C mass transfer \citep{Lauterborn1970}, by contrast, occurs during or after helium shell burning and is often dynamically unstable, ejecting the donor's hydrogen envelope into the circumstellar environment rather than transferring it cleanly onto the companion. If this episode occurs sufficiently close to core collapse, the SN ejecta may collide with the resulting circumstellar material (CSM) within a few years after explosion.

In addition to late Case~C mass transfer, at least two other mechanisms can produce eruptive mass ejection close to core collapse. First, vigorous convection during the final nuclear burning stages can excite gravity waves that transport energy to the stellar envelope, driving mass outbursts on timescales of years to decades \citep{Quataert12, Fuller17, Fuller18}. Second, stripped helium stars can re-expand during these same late burning stages \citep{Laplace20}, potentially re-initiating binary mass transfer timed by the nuclear burning clock rather than the companion's evolution. Both mechanisms naturally produce dense CSM close to the star at the time of explosion, detectable through late-time ejecta--CSM interaction.

Such ejecta--CSM interaction produces UV-bright emission through Mg~II and Fe~II lines \citep{ChevalierFransson2017}, as demonstrated by the Type~Ia--CSM SN~2015cp \citep{GrahamM2019}. A late-time UV imaging survey can therefore directly probe the mass-loss histories of SE~SNe by searching for this interaction signature. The Zwicky Transient Facility (ZTF; \citealp{Bellm2019,Graham2019,Dekany2020}) Bright Transient Survey (BTS; \citealp{Fremling2019rcf,Perley2019b}) provides a controlled sample of SE SNe suitable for this purpose.

Late-time CSM interaction has already been observed in several SE~SNe. Photometry and spectra of the Type~IIb SN~1993J taken years after explosion show a broad-band excess and hydrogen features requiring CSM interaction \citep{Matheson2000}. SN~2013df \citep{Maeda2015} and SN~2018mpl \citep{Fremling_aalrxas} display similar signatures. All three SNe~IIb also exhibit exceptionally strong early-time cooling phases, which may themselves also indicate the presence of nearby CSM \citep{Maeda2015}. This is somewhat in tension with standard binary models for SE~SNe, which tend to favor Case~A (SNe~Ib, Ic; \citealp{Yoon2010}) or Case~B (SNe~IIb; \citealp{Yoon2017,Sravan2019}) systems with long delays between mass transfer and explosion, producing mass-loss rates close to the time of explosion far too low to explain the interaction seen in SN~1993J and analogs.

\begin{figure*}[t!]
\centering
\includegraphics[width=0.95\linewidth]{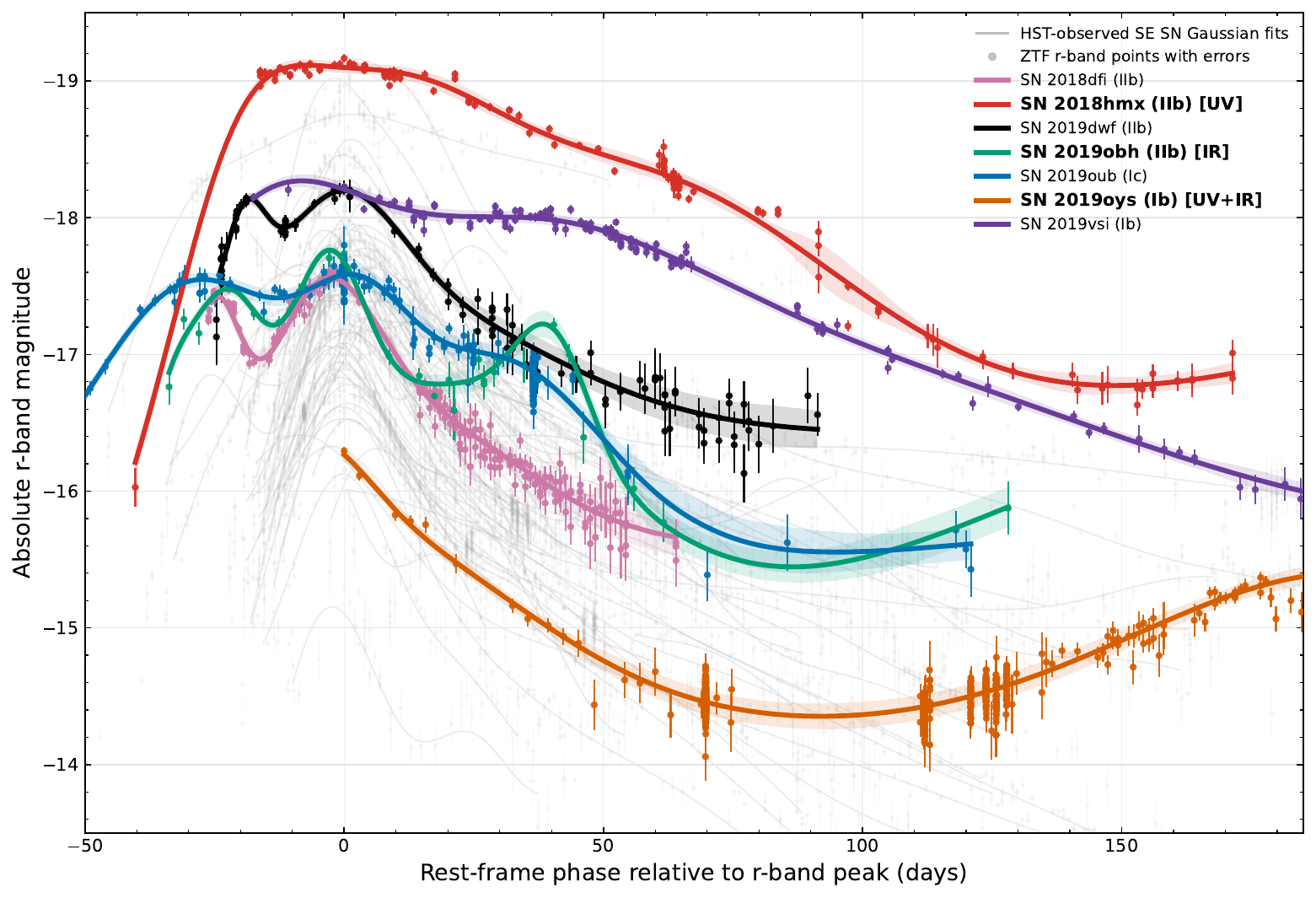}
\caption{Rest-frame absolute ZTF forced-photometry $r$-band light curves for the HST-observed SE~SN sample. The gray and colored curves show the full HST-observed SE~SN sample. Colored curves highlight objects with long-lived, multi-peaked, or unusually luminous optical evolution: SN~2019oys (ZTF19abucwzt; Ib; orange), SN~2019obh (ZTF19abqykei; IIb; green), SN~2019oub (ZTF19abtsnyy; Ic; blue), SN~2018dfi (ZTF18abffyqp; IIb; pink), SN~2019dwf (ZTF19aarfkch; IIb; black), SN~2018hmx (ZTF18accnoli; IIb; red), and SN~2019vsi (ZTF19acxpuql; Ib; purple). Boldface legend entries indicate HST UV detection or a strong late-time WISE excess, or both; the bracketed legend tags specify [UV], [IR], or [UV+IR]. SN~2019oys is detected in both HST UV and WISE, SN~2019obh is detected in IR only, SN~2018hmx is detected in UV only, and the remaining four highlighted outliers show neither signature.} 

\label{fig:lc_spec}
\end{figure*}

Strong cooling phases have now been identified across all SE~SN subtypes, including Type~Ic iPTF15dtg \citep{Taddia2016} and Type~Ic iPTF14gqr \citep{De2018}, and Type~Ib SN~2019dge \citep{Yao2020}. Among the SE~SNe selected for and observed by our HST program, we find a subset with unusual optical behavior (strong cooling signatures, slow evolution, or multiple peaks) in all three main subtypes (Fig.~\ref{fig:lc_spec}; the sample selection and photometry are described in Sects.~\ref{sec:sample}--\ref{sec:observations}). The bulk of the sample follows the narrower locus expected from canonical SE~SN light-curve evolution (e.g., \citealp{Taddia2015}), but the outliers hint at an additional power source beyond \Nif\ decay: either CSM interaction or a central engine \citep{Kasen2010,Woosley2010}.

The highlighted SE-SN light curves in Fig.~\ref{fig:lc_spec} are qualitatively similar to SE~SNe in which late-time CSM interaction has been confirmed spectroscopically, such as the Type~Ib$\to$IIn transitional SN~2014C \citep{Milisavljecic2015}, SN~2019oys \citep{Sollerman2020}, and the luminous Type~Ic-BL SN~2017ens which evolved into a IIn-like phase \citep{Chen2018_2017ens}.
However, optical light-curve morphology alone is not decisive: broad or structured light curves can also arise from central-engine input or simply from unusually massive ejecta \citep{Valenti:2012aa,Taddia2016,2023Karamehmetoglu}.

This ambiguity motivates a multi-wavelength approach. Late-time flux at SN sites can arise from radioactive decay, ongoing ejecta--CSM interaction, central-engine power, light echoes, or unrelated stellar populations \citep{BaerWay2024}. Secure UV detections of core-collapse SNe are rare, and host emission can easily mimic them \citep{Inkenhaag2025}. The primary measurement in this paper is the late-time HST UV detection rate in a statistical SE~SN sample drawn from BTS. ZTF light curves are used to define phases and provide optical context for the individual objects, while overlapping Wide-field Infrared Survey Explorer (WISE; \citealp{Wright2010}) W1/W2 light curves provide a complementary mid-infrared check on dust-heated interaction where available \citep{Mo2025,Myers2024}. All magnitudes in this paper are reported in the AB system.

The rest of this paper is organized as follows. In Sect.~\ref{sec:sample} we describe the survey sample and its selection. In Sect.~\ref{sec:observations} we present the HST observations, the source detection and classification procedure, and the late-time WISE analysis. In Sect.~\ref{sec:results} we report the UV and mid-IR detections and describe each object with a secure counterpart. In Sect.~\ref{sec:analysis} we use a physics-based CSM interaction model and Bayesian MCMC inference to constrain the late-time CSM interaction rate among SE~SNe. We discuss the results and present our conclusions in Sect.~\ref{sec:conclusions}.

\section{Survey Sample}\label{sec:sample}
Targets for our HST Cycle~29 Snapshot program (SNAP-16657; \citealp{Fremling21_SNAP16657}) were drawn primarily from the untargeted ZTF Bright Transient Survey (BTS)\footnote{\url{https://sites.astro.caltech.edu/ztf/bts/bts.php}} \citep{Fremling2019rcf,Perley2019b,Rehemtulla2024}.
The HST program contained 163 targets, of which 121 were ultimately observed. Based on current BTS classifications retrieved via the Fritz marshal \citep{vanderWalt2019,Coughlin2023}, the 121 executed observations break down as follows: \Ntotal\ SE~SNe (27~SNe~Ib, 30~SNe~IIb, 27~SNe~Ic, and 7~SNe~Ic-BL), \Ncomparison\ interacting Type~II control object which we retain because of prior spectroscopic evidence for very strong late-time CSM interaction, 
\NtotalSLSN\ SLSNe (20~SLSNe-I and 1~SLSN-I.5) not analyzed in this paper, and \Nexcluded\ objects excluded from the analysis in this paper because their updated classifications place them outside the normal SE~SN types (5~SNe~Ibn, 1~SN~Ia-CSM, 1~SN~Ia, and 1 additional Type~II SN). 

The basic properties of the \Ntotal\ SE~SNe together with the retained Type~II comparison object are listed in Table~\ref{tab:targets}. 
The SE~SN sample spans $z = \MinRedshift$--$\MaxRedshift$ (median $z = \MedianRedshift$), and the HST observations span \MinPhase--\MaxPhase\ days after first detection (Sect.~\ref{sec:observations}), targeting the phase range in which radioactive nickel powered SE-SN emission should be faint while delayed interaction from recently ejected shells could still remain UV bright. Overlapping WISE coverage adds a complementary test for dust-heated late-time emission over much of the same phase range. Because the program used Snapshot time, the executed sample is necessarily an opportunistic subset of the original list of 163 targets for the program, but that trade-off allowed us to survey a much larger number of late-time UV targets than would otherwise have been possible.

\subsection{Sample selection}
The HST proposal SNAP-16657 \citep{Fremling21_SNAP16657} was designed to test how often delayed-onset CSM interaction appears in hydrogen-poor core-collapse SNe during the years after explosion. Targets were drawn from three groups. The first group consisted of ``normal'' stripped-envelope SNe from BTS. Starting from the 327 SE~SNe known to BTS in March 2021, we required (i) a measurable offset from the host-galaxy nucleus to limit host UV contamination, (ii) well-sampled $g_\mathrm{ZTF}$ and $r_\mathrm{ZTF}$ light curves so that the explosion epoch and basic light-curve properties could be estimated, (iii) low Galactic reddening, $E(B-V)_\mathrm{MW}<0.15$~mag, and (iv) an expected HST phase of roughly 200--1200 days during the Cycle 29 window. These practical quality and visibility cuts yielded 80 targets, including SNe Ic-BL.

The second group consisted of SE~SNe with optical light curves that could plausibly signal strong pre-SN mass loss or ongoing interaction. In practice this meant long rise times ($>30$~days), unusually luminous peaks ($M_{\rm peak}<-19$~mag), or clear multiple light-curve peaks \citep{Sharma2025}. For these rarer targets we kept the same requirements on host offset, light-curve quality, and phase, but relaxed the Galactic-extinction cut. This added 39 targets. The proposal also included a separate SLSN-I component, which explains part of the executed 121-object HST sample but these are not analyzed further here. 

Of the \Ntotal\ observed SE~SNe, 14 meet the criteria above used to select unusual SNe. To test whether their inclusion biases the sample, we compared the distributions of peak luminosity, light-curve width, and time spent brighter than $M=-17$ against a 384-object BTS control sample\footnote{Here the control sample is defined as the non-HST BTS SE~SN sample discovered between 2018 March 22 and 2024 January 2 for which the same smoothed ZTF $r$-band light curves could be constructed from ZTF forced-photometry measurements.} and find no significant enrichment in luminous, slow-rising, or long-lived outliers. The random target execution of the HST Snapshot program therefore produced an apparently unbiased SE~SN sample in terms of optical light-curve properties, and we analyze all \Ntotal\ SE~SNe together without separating canonical and optically unusual objects.

\begin{figure*}
    \centering
    \includegraphics[width=1.0\linewidth,page=1]{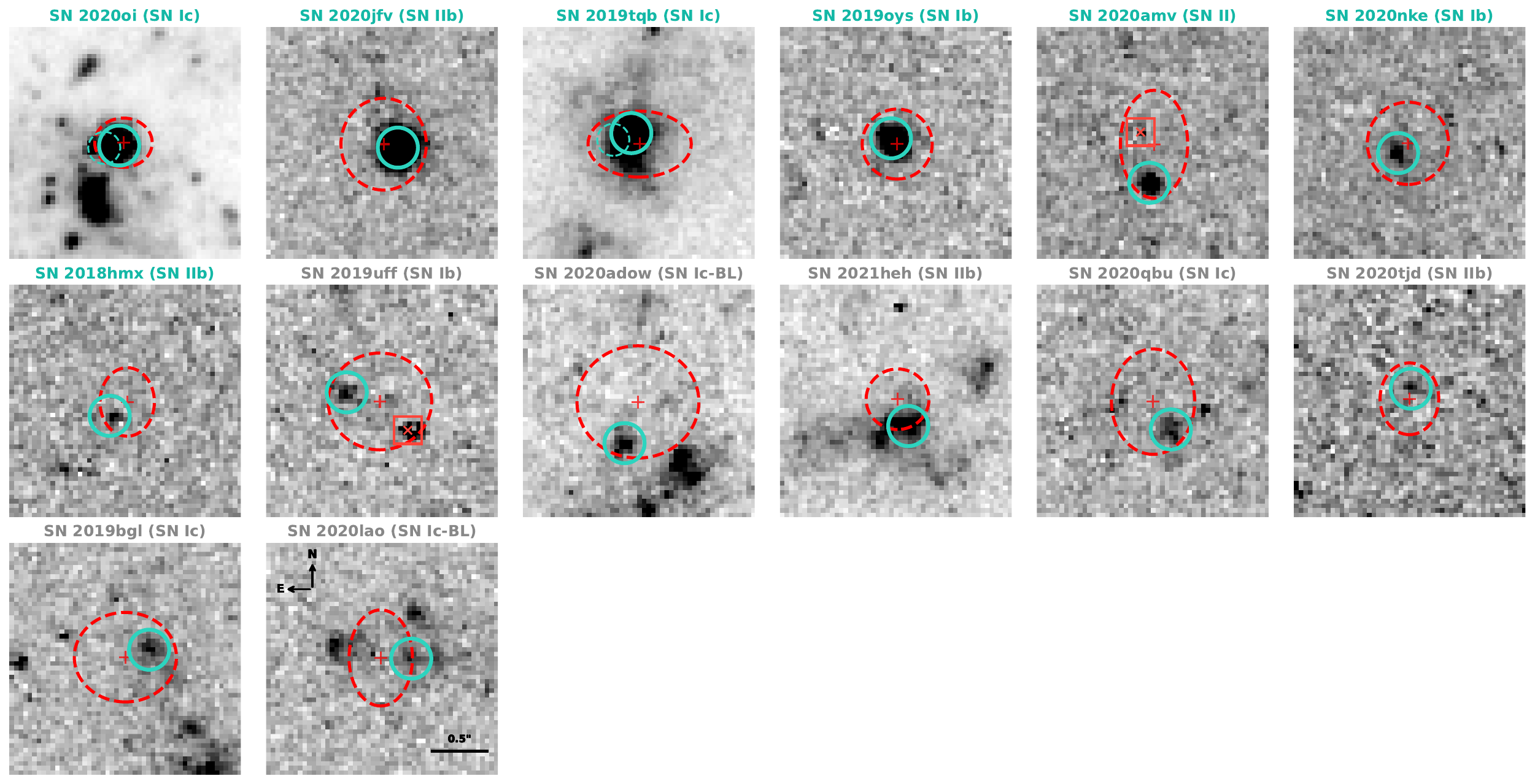}
    \caption{HST WFC3/UVIS cutouts for all candidate detections (confirmed and ambiguous). Each panel shows a $1\farcs0$ radius region centered on the expected SN position. Red dashed ellipses mark the $3.5\sigma$ astrometric uncertainty region derived from the per-axis ZTF and HST WCS uncertainties (Table~\ref{tab:astrometry}). A thick solid cyan circle marks the adopted primary counterpart used for the reported photometry, while thinner dashed cyan circles mark additional FLC-confirmed sources within the uncertainty region. Red squares mark sources rejected by the FLC comparison as cosmic rays. The panel-title color indicates whether the retained source is classified as secure (cyan) or ambiguous (gray). The grayscale is inverted (black = positive flux).}
    \label{fig:det_cutouts}
\end{figure*}

\begin{figure*}
    \centering
    \includegraphics[width=1.0\linewidth,page=1]{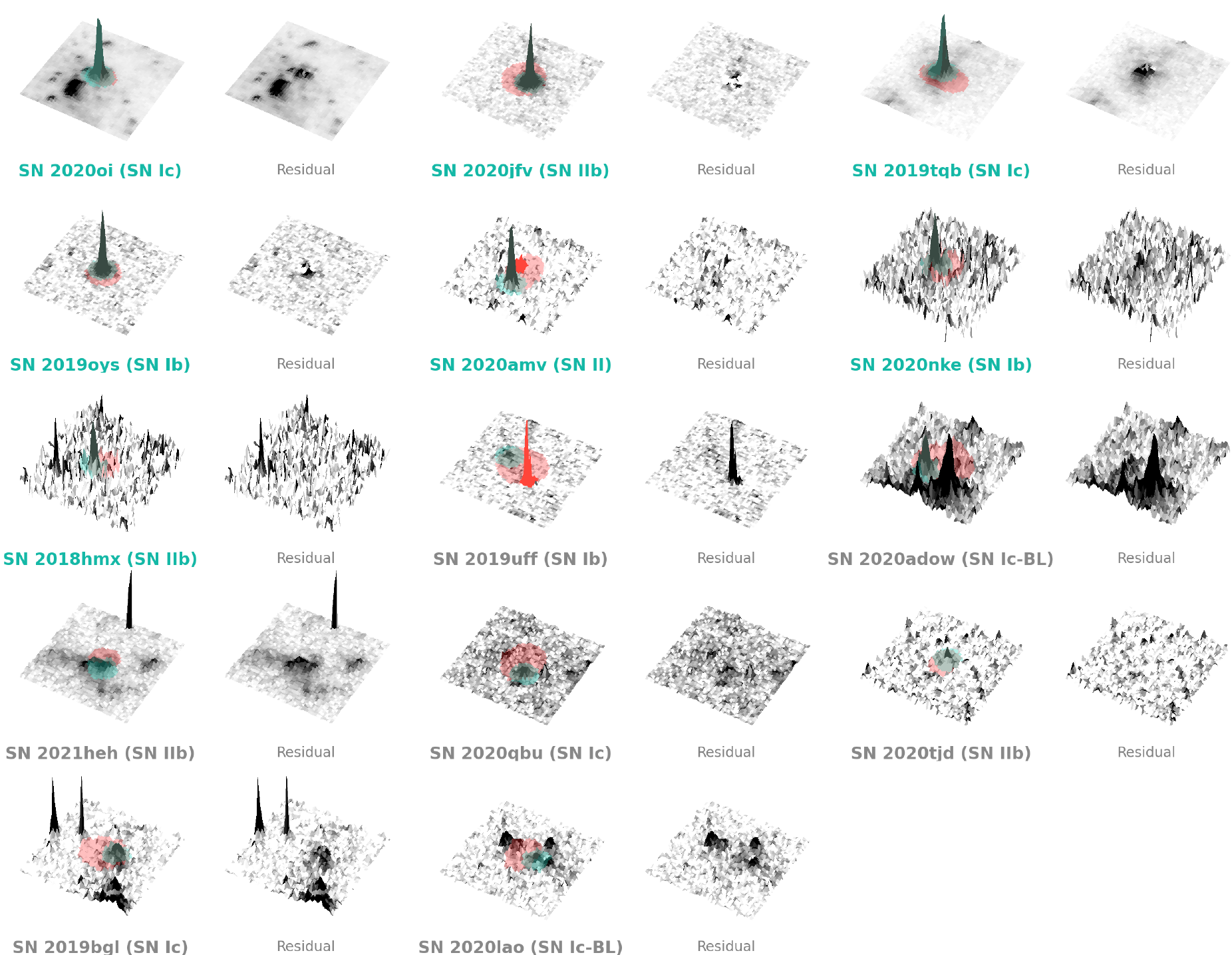}
    \caption{Three-dimensional surface representations of HST WFC3/UVIS cutouts for all candidate detections (confirmed and ambiguous). Each panel pair shows a $1\farcs0$ radius region with the DRC image (left) and the PSF-subtracted residual (right) as peak-normalized flux surfaces. Transparent red shading marks the $3.5\sigma$ astrometric uncertainty ellipse projected onto the image surface. Transparent cyan shading marks the primary candidate position. Solid red patches mark sources rejected as cosmic rays by the FLC comparison. For secure detections, the residual panels demonstrate the quality of the PSF fit: a clean subtraction produces a flat residual surface at the candidate position, confirming the point-source nature of the detection. The panel-title color indicates whether the retained source is classified as secure (cyan) or ambiguous (gray).}
    \label{fig:det_cutouts_3d}
\end{figure*}

\section{Observations and data reduction}\label{sec:observations}

\subsection{HST observations}\label{sec:hst_obs}
Our late-time interaction search combines ZTF optical light curves, HST near-UV imaging, and WISE mid-IR forced-photometry light curves. Each HST visit consisted of two dithered\footnote{We use the UVIS-LINE-DITHER pattern with a point spacing of $0\farcs145$ and pattern orientation of $46\fdg84$ with the target placement UVIS2-C1K1C-CTE.} exposures. The observing and reduction strategy was built around a simple question for each target: does the late-time UV image contain a point source consistent with ongoing interaction, or only host-galaxy structure at the SN position? The proposal assigned filters by redshift: low-redshift targets ($z < 0.15$) were observed in F275W ($\lambda_{\rm pivot} \approx 2704$~\AA) with two 600~s sub-exposures for a total of 1200~s, while higher-redshift targets were observed in F336W ($\lambda_{\rm pivot} \approx 3355$~\AA) with two 800~s sub-exposures for a total of 1600~s. Because the entire SE~SN sample lies below $z = 0.15$ (Sect.~\ref{sec:sample}), all \Ntotal\ SE~SN observations and the retained Type~II comparison object we study in this paper were observed with the F275W filter. Example HST image cutouts are shown in Figures~\ref{fig:det_cutouts} and \ref{fig:det_cutouts_3d}.

All photometry is corrected for Milky Way foreground extinction using the \citet{Schlafly11} recalibration of the \citet{Schlegel98} dust maps, queried via the IRSA Galactic Dust Reddening and Extinction Service\footnote{\url{https://irsa.ipac.caltech.edu/applications/DUST/}}. We adopt the \citet{Fitzpatrick99} $R_V=3.1$ extinction curve, which gives $A_{\rm F275W}/E(B-V)_\mathrm{MW} = 6.47$ at the F275W pivot wavelength. The median foreground extinction for the SE~SN sample is $A_{\rm F275W} = 0.21$~mag; one target (SN~2019oub) has $A_{\rm F275W} = 1.99$~mag due to its low Galactic latitude. Per-object $E(B-V)_\mathrm{MW}$ values are listed in Table~\ref{tab:targets}.

\subsection{Source detection and candidate classification}\label{sec:classification}
We searched each drizzled\footnote{DRC (drizzle-combined) images are the final co-added products delivered by the HST WFC3 calibration pipeline, combining dithered sub-exposures into a single geometrically corrected frame. FLC (flat-fielded, CTE-corrected, calibrated) images are the individual CTE-corrected sub-exposures before co-addition.} (DRC) HST image with a multiscale \texttt{DAOStarFinder} run and retained only sources that fell inside the $3.5\sigma$ astrometric error ellipse. We use $3.5\sigma$ rather than the conventional $3\sigma$ because the enclosed probability of a two-dimensional Gaussian within radius $r\sigma$ is $1-\exp(-r^2/2)$: at $3\sigma$ this is only 98.9\%, implying ${\sim}1.1$ missed true counterparts per 100 targets, whereas at $3.5\sigma$ the enclosed probability rises to 99.8\%, reducing the expected misses to ${\sim}0.2$ per 100 targets. Each candidate was then projected back to the two individual (FLC) sub-exposures. The minimum requirement for a real source is a recovery at $\geq3\sigma$ in both FLC frames; one-frame-only events are flagged as likely cosmic rays and rejected. When more than one FLC-confirmed source lies inside the ellipse, we adopt as the primary counterpart the one with the smallest ellipse-normalized offset, $r_{\rm ell}^2=(\Delta x/a)^2+(\Delta y/b)^2$, where $a$ and $b$ are the RA and DEC semi-axes of the $3.5\sigma$ search ellipse.

For each candidate we measure PSF photometry using the STScI empirical spatially variable WFC3/UVIS PSF library \citep{Anderson2006}, which provides 56 focus-diverse realizations per filter on a $7\times8$ grid across the detector, each stored as a $101\times101$ pixel model at $4\times$ the native pixel scale. The PSF model is aligned to the data with subpixel precision via DFT registration \citep{Guizar2008}. The quality of the PSF subtraction is illustrated in the 3D surface cutouts of Figure~\ref{fig:det_cutouts_3d}, where the residual panels show that confirmed detections leave clean, flat residuals at the candidate position after PSF removal, confirming their point-source nature.

To quantify host confusion, we also collect control sources in the same DRC image outside the astrometric ellipse but within a $5\farcs0$ radius of the SN position. These control sources define both the local source surface density in the field and a pooled survey-wide luminosity distribution of compact host features. For the adopted primary candidate we record $p(\mathrm{HII})$, defined here as the fraction of pooled control sources with luminosities at least as large as the candidate luminosity. This quantity is therefore an empirical host-contamination metric, not a literal posterior probability that the source is an H~II region. We also record the candidate brightness relative to the local control distribution for diagnostic purposes, but do not use that quantity as a hard threshold in the final classification.

Before rival counting, we require every candidate to have a PSF goodness-of-fit $R^{2}\geq0.60$; sources below this threshold are rejected as non-pointlike detections that cannot be reliably distinguished from background fluctuations or extended host features. A target remains in the sample as an ambiguous candidate rather than a secure detection if any one of three conditions is met: (i)~a second FLC-confirmed source with $R^{2}\geq0.60$ has a positional likelihood at least 50\% of the primary candidate's, (ii)~the adopted primary source is flagged as non-pointlike, requiring both a PSF-fit $R^{2}<0.70$ and either a measured FWHM exceeding 1.6 times the nominal PSF FWHM or an absolute roundness greater than 1.0, or (iii)~$p(\mathrm{HII})>0.5$. Condition~(iii) is a host-confusion downgrade: it moves a candidate from secure to ambiguous but does not convert it into a non-detection or remove it from the sample. We also record the Poisson expectation value of unrelated control sources inside the astrometric ellipse, $N_{\rm exp}$, defined as the local control-source surface density multiplied by the ellipse area, as a diagnostic reported in Table~\ref{tab:candidate_logic}. All reported UV magnitudes for secure detections are PSF fits at the adopted primary-candidate position; ambiguous cases are shown in the survey cutouts (Figures~\ref{fig:all_cutouts} and \ref{fig:det_cutouts_3d}) but are not assigned unique UV magnitudes in the tables. The astrometric search region itself is defined by a $3.5\sigma$ error ellipse whose semi-axes in RA and DEC are computed from four independent terms added in quadrature: the ZTF per-axis systematic uncertainty ($0\farcs06$; \citealp{Masci2019}), the statistical centroid uncertainty of the filtered ZTF alert positions for each target, the HST WCS uncertainty measured empirically from Gaia~DR3 cross-match residuals, and the guide-star catalog alignment uncertainty. The full per-object astrometric uncertainty budget is given in Table~\ref{tab:astrometry}.

\subsection{Limiting magnitudes}\label{sec:limits}
For non-detections, a simple PSF-based $3\sigma$ noise estimate is computed by fitting the PSF model at ${\sim}50$ random positions at varying radii around the SN site, sigma-clipping the resulting flux distribution, and taking three times its standard deviation. This estimate is retained only as a reference diagnostic and is not the quantity used in the figures, tables, or CSM modeling. Instead, we measure astrometry-aware limiting magnitudes by sampling plausible SN positions from the same two-dimensional Gaussian astrometric posterior used to define the $3.5\sigma$ ellipse, truncated at that ellipse. At each sampled position we inject a normalized point-source PSF into the DRC image and both FLC sub-exposures, rerun the DRC source finder, and require recovery in the DRC image and at $\geq3\sigma$ in both FLC frames. Repeating this over the sampled astrometric posterior yields a required recovery magnitude at each position and therefore a cumulative coverage curve for that target. We adopt a 70\% injection/recovery limit, which is the magnitude at which a source would be recovered over 70\% of the sampled astrometric posterior, as the single-number upper limit reported in the tables, shown in the luminosity-phase figure, and used in the MCMC. This threshold balances depth against astrometric uncertainty: a 50\% limit would reflect the depth at the most likely SN position but would be optimistic for targets whose true position may lie in a more confused part of the ellipse, while a 90\% limit is driven by the hardest tail positions and is unnecessarily conservative for most targets.

For the SE~SN sample analyzed below, this procedure yields \Nconfirmed\ secure UV counterparts and \Nambiguous\ additional ambiguous in-ellipse candidates. In simple terms, our overall analysis indicates that sources fainter than $M=-10$ cannot be cleanly separated from host-galaxy background without reference-image subtraction, so we treat those cases as ambiguous.

\subsection{WISE mid-infrared light curves}\label{sec:wise_obs}
We also inspected local W1 and W2 difference-image light curves as an independent tracer of dust-heated interaction power, following the same forced-photometry and cutout-based approach used in recent delayed-interaction mid-IR studies \citep{Myers2024,Mo2025}. 
For each target we converted the forced photometry to mJy using the tabulated WISE zero points, measured a pre-SN baseline from epochs more than 30~days before optical peak, and searched for positive residuals at phases $>300$~days. Because subtraction artifacts around bright host nuclei often produce both positive and negative excursions, we folded the pre-SN scatter into the late-time significance estimate and required any claimed excess to exceed the largest positive pre-SN excursion in the same band. This baseline-based cut is intentionally conservative and rejects several visually suggestive but low-significance cases. For the WISE detections that pass these criteria, we also inspected the difference-image cutouts from the forced-photometry pipeline, which confirmed the authenticity of the detections.

\begin{figure*}
    \centering
    \includegraphics[width=1.0\linewidth]{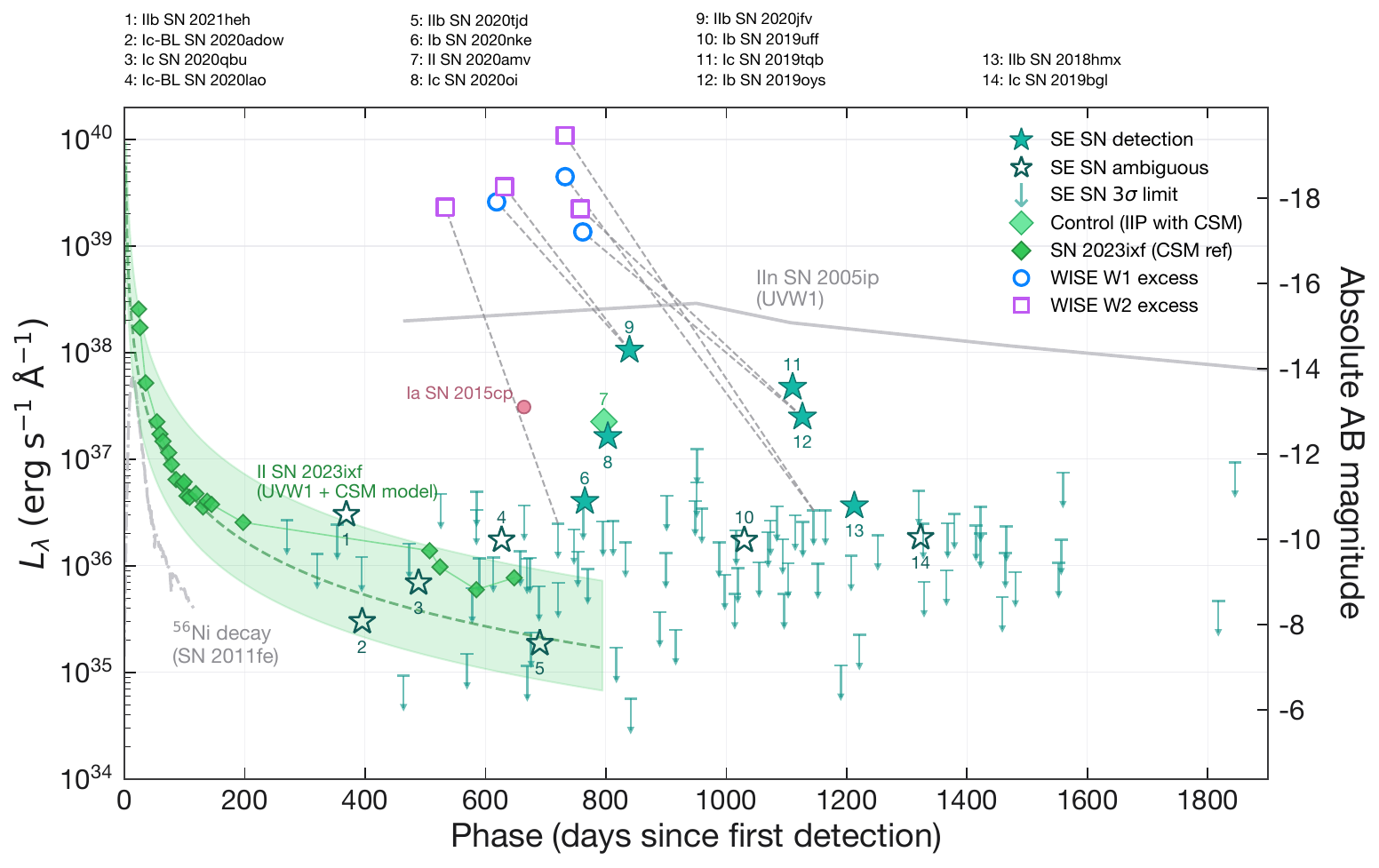}
    \caption{Luminosity density $L_\lambda$ (erg~s$^{-1}$~\AA$^{-1}$) at each filter's pivot wavelength vs.\ phase (from first ZTF 5$\sigma$ forced-photometry detection) for our HST survey. The SE~SN sample is observed in F275W ($\lambda_\mathrm{pivot} \approx 2704$~\AA); reference models, SN~2023ixf, and WISE data are plotted at their respective wavelengths. Symbols are defined in the legend; numbered labels identify individual detections. Downward arrows use the adopted $70\%$-coverage astrometry-marginalized PSF injection/recovery limits. Open circles/squares connected by dashed lines show mean late-time WISE W1/W2 excesses (read against the right-hand AB-magnitude axis). Gray curves show reference models: $^{56}$Ni decay of SN~2011fe in UVW1 (dot-dashed), Type~IIn SN~2005ip in UVW1 (solid), and the Type~Ia--CSM SN~2015cp at 664~d (red circle; \citealp{GrahamM2019}). Green diamonds trace the Swift/UVOT UVW1 light curve of SN~2023ixf from $\sim$24 to $\sim$648~d \citep{Jacobson-Galan25}; the green shaded band shows our CSM interaction model (Section~\ref{sec:csmmodel}) for $M_\mathrm{CSM} = 0.005$--$0.03~M_\odot$. Our SE~SN sample yields \Nconfirmed\ secure and \Nambiguous\ ambiguous late-time UV detections.}
    \label{fig:results}
\end{figure*}

\section{Results}\label{sec:results}

Figure~\ref{fig:results} and Table~\ref{tab:targets} summarize the HST UV measurements for all \NtotalPaper\ targets. Among the \Ntotal\ SE~SNe we retain \Nconfirmed\ secure and \Nambiguous\ ambiguous late-time UV counterparts, corresponding to a secure detection rate of $\Nconfirmed/\Ntotal \approx \ConfirmedRate\%$ (or $\DetectionRate\%$ when ambiguous cases are included). We constrain the underlying CSM interaction rate with a Bayesian MCMC analysis in Sect.~\ref{sec:mcmc}. The secure SE~SN detections span SNe~Ib, IIb, and Ic; the retained Type~II comparison object SN~2020amv is also securely detected. The secure counterparts have absolute magnitudes in the range $M_{\rm UV}\sim-10$ to $-14$. In the mid-IR, four SE~SNe show robust late-time W1/W2 excesses: SN~2019obh, SN~2019oys, SN~2020jfv, and SN~2020sgf (Table~\ref{tab:wise}). No additional SE~SN remains even tentatively significant after accounting for pre-SN scatter; SN~2019bao and SN~2020amv show positive late-time residuals at only the 2--4$\sigma$ level and are not counted as secure IR detections. Brief notes on every secure or ambiguous UV candidate are collected in Appendix~\ref{sec:candidate_notes}, and the aligned DRC/FLC diagnostics are shown in Figures~\ref{fig:diag_candidates} and \ref{fig:diag_wise}. 
Below we describe each object with a secure UV counterpart or a robust late-time WISE excess.

\paragraph{SN 2018hmx (ZTF18accnoli).}
A Type~IIb at $z=0.038$, classified as SN~II on the TNS \citep{Fremling2018_2018hmx} based on early spectra showing broad H$\alpha$ with an unusually high expansion velocity of ${\sim}17{,}000$~km~s$^{-1}$. We adopt a SN~IIb classification because subsequent SEDM spectra are very well matched to SN~1993J. The optical light curve declines more slowly than most SE~SNe, but still consistent with the BTS SE SN sample, which suggests it may be a transitional IIb/II with a somewhat more massive residual hydrogen envelope than average SNe IIb, or that its early light curve is already affected by CSM interaction. Our HST image, obtained 1213~days after first detection, contains a single FLC-confirmed source inside the astrometric ellipse with no competing counterpart. The local background is structured but the association is secure. The measured UV magnitude is $m_{\rm F275W}=25.71$, corresponding to $M_{\rm UV}\approx-10.8$. No dedicated multi-wavelength study of SN~2018hmx has been published to date.

\paragraph{SN 2019obh (ZTF19abqykei).}
A Type~IIb \citep{Fraser2019_2019obh} in one of the spiral arms of MRK~0618. Its ZTF light curve is structured and unusually broad for the subtype, remaining detectable for about 80~days. We do not detect a UV source in the HST image obtained roughly 1100~days after first detection, but NEOWISE shows a clear long-lived IR excess that starts before the ZTF discovery and continues to the end of the IR data set, about 1000~days after the last optical detection. No dedicated multi-wavelength study of SN~2019obh has been published to date.

\paragraph{SN 2019oys (ZTF19abucwzt).}
This Type~Ib \citep{Fremling2019_2019oys} is a case in which late-time spectra revealed ongoing CSM interaction \citep{Sollerman_2020}; radio data support the same picture \citep{Sfaradi_2024}. The optical light curve developed a pronounced second bump and remained active for more than 1400~days after the first ZTF detection. Our HST image was obtained 858~days after the second optical peak, while the source was still visible in the optical, and we detect a UV counterpart at the SN position. NEOWISE also shows a strong IR excess that rises near optical maximum and peaks around 900~days after the optical peak, making SN~2019oys one of the clearest interaction cases in the sample.

\paragraph{SN 2019tqb (ZTF19acjtpqd).}
A Type~Ic at $z=0.017$ \citep{Dahiwale2020_2019tqb}. Our HST image, obtained ${\sim}1110$~days after first detection, contains two FLC-confirmed sources inside the astrometric ellipse, but the secondary source has a PSF goodness-of-fit $R^{2} = 0.17$, far below the $R^{2}\geq0.60$ threshold (Sect.~\ref{sec:classification}), and is therefore rejected as a viable rival. The primary candidate ($R^{2} = 1.00$) is compact and well-centered in the ellipse, making this a secure late-time UV detection. We do not find a corresponding WISE excess.

\paragraph{SN 2020oi (ZTF20aaelulu).}
A nearby Type~Ic located close to the center of M100. It is one of the best-studied SE~SNe at early times: early radio observations revealed a non-equipartition shock propagating through dense CSM \citep{Horesh2020}, near-infrared spectroscopy provided the first detection of CO and dust formation in a Type~Ic SN \citep{Rho2021}, and ALMA band~3 data showed that the CSM deviates from a smooth wind profile within ${\sim}10^{15}$~cm, tracing mass-loss fluctuations in the progenitor's final months \citep{Maeda2021}. An early-time UV/optical excess was also attributed to CSM interaction \citep{Gagliano_2022}. The optical light curve rose rapidly, peaked within 11~days, and then declined for about 100~days with modest structure. The late-time HST image contains two FLC-confirmed in-ellipse sources, but one is clearly preferred by the astrometric ranking and we therefore retain a secure primary counterpart. The field remains strongly host contaminated, so the existence of a real late-time UV source is secure whereas the interpretation of that flux as delayed interaction is still somewhat uncertain. We do not find a corresponding IR excess.


\paragraph{SN 2020amv (ZTF20aahbamv).}
This is not a stripped-envelope SN, but we retained it as a comparison object because its optical spectra already showed strong $H\alpha$ emission and clear evidence for CSM interaction \citep{Sollerman_2021} when we were building the sample for SNAP-16657. The light curve rose and declined quickly before entering a plateau lasting ${\sim}250$~days, with the SN remaining detectable for ${\sim}400$~days after peak. With our revised astrometric solution, the HST image contains a single FLC-confirmed UV source inside the current error ellipse, and it serves as a useful comparison detection for the classification procedure. We do not see a comparably clear IR excess.

\paragraph{SN 2020jfv (ZTF20abgbuly).}
First detected by ATLAS \citep{Tonry2020_2020jfv}. Its light curve remained active for more than 2000~days and developed a second peak about 800~days after discovery. A SEDM \citep{SEDM2018,Rigault2019} spectrum taken at ${\sim}50$~d after explosion shows excellent agreement with SN~IIb spectra, which lead to this SN being classified as a SN~IIb \citep{Dahiwale2020_2020jfv}. Later spectra show nebular H$\mathrm{\alpha}$ emerging, and indicate a strong resemblance to interacting Type~IIn SNe \citep{Sollerman_2021}. Our HST image was obtained during the second optical peak and shows a significant UV detection at the SN position. NEOWISE reveals a contemporaneous IR rise that begins near optical maximum and continues through the end of the data. We therefore regard SN~2020jfv as a confident interaction detection.

\paragraph{SN 2020sgf (ZTF20abxpoxd).}
A SN~Ib in the edge-on galaxy UGC~02700, initially classified as a SN~Ic from a low-resolution SEDM spectrum \citep{Dahiwale2020_2020sgf}, but later higher-resolution BTS follow-up spectra show multiple He I absorption features (Qin et al., in prep.). SN~2020sgf was selected because of its unusual optical light-curve evolution: a rapid rise, a structured peak, and a nearly flat decline lasting  ${\sim}450$~days before becoming undetectable by ZTF. We do not detect UV emission in the HST image taken ${\sim}678$~days after peak light. NEOWISE, however, shows a clear late-time W2 excess (${\sim}17\sigma$) starting within days of the first optical detection and continuing for more than 500~days after optical peak, including the epoch of the HST observation. The W1 band does not reach our strong-detection threshold, but this is attributable to the ${\sim}2.6\times$ higher difference-image noise at 3.4~$\mu$m from the bright edge-on host; the measured W1 flux ($M_{W1} \approx -17.4$) is consistent with a source of comparable brightness to the W2 detection ($M_{W2} \approx -17.8$). We therefore classify SN~2020sgf as an IR-selected interaction candidate. No dedicated study of SN~2020sgf has been published to date.

\section{Analysis}\label{sec:analysis}

\subsection{Constraining the CSM interaction rate}
\label{sec:mcmc}

To constrain the fraction of SE~SNe that undergo late-time CSM interaction, we performed a Bayesian MCMC analysis using a physics-based CSM interaction model and the \Ntotal\ SE~SNe in our sample (using both detections and non-detections, treating ambiguous cases as non-detections). The inferred $f_\mathrm{CSM}$ should be interpreted as the fraction of SE~SNe that produce UV-bright delayed interaction within the phase and luminosity window probed by this survey.

\subsubsection{Physics-based CSM model}\label{sec:csmmodel}

We adopt a semi-analytical energy-budget model for UV emission from SN ejecta colliding with a circumstellar shell, following the framework of \citet{Chevalier17} and \citet{Chatzopoulos12}. The model assumes a CSM shell of mass $M_\mathrm{CSM}$ located at radius $R_\mathrm{CSM} = v_\mathrm{ej} \times t_\mathrm{onset}$, where $t_\mathrm{onset}$ is the time after explosion at which the fastest ejecta reach the shell. The shell has a radial thickness $\Delta R = f_\mathrm{thick} \times R_\mathrm{CSM}$, where the thickness fraction $f_\mathrm{thick}$ is treated as a free parameter. Physically, $f_\mathrm{thick}$ encodes the duration of the pre-explosion mass-ejection episode: a thicker shell corresponds to longer-lived but fainter emission, since the shock power is spread over a longer crossing time. The key physical quantities are:
\begin{itemize}
\item The shock velocity $v_\mathrm{sh} = v_\mathrm{ej}\, M_\mathrm{ej} / (M_\mathrm{ej} + M_\mathrm{CSM})$ from momentum conservation.
\item The deposited kinetic energy $E_\mathrm{kin} = \frac{1}{2}\, M_\mathrm{CSM}\, v_\mathrm{sh}^2$.
\item The emission timescale $t_\mathrm{em} = \max(t_\mathrm{cross},\, t_\mathrm{diff})$, where $t_\mathrm{cross} = \Delta R / v_\mathrm{sh}$ is the shell crossing time and $t_\mathrm{diff} = \sqrt{\kappa\, M_\mathrm{CSM} / (4\pi\, c\, v_\mathrm{sh})}$ is the photon diffusion time.
\item The peak luminosity $L_\mathrm{peak} = \epsilon_\mathrm{rad}\, E_\mathrm{kin} / t_\mathrm{em}$, where $\epsilon_\mathrm{rad} = 0.1$ is the radiative efficiency.
\end{itemize}
The light curve rises linearly over $\min(t_\mathrm{diff},\, t_\mathrm{em}/5)$, remains at $L_\mathrm{peak}$ for $t_\mathrm{em}$ (sustained shock crossing), and then declines as $t^{-5/3}$ as the shocked material cools \citep{Chevalier82}. The fraction of bolometric luminosity in the F275W bandpass is fixed at $f_\mathrm{UV} = 0.3$.\footnote{Radiation-hydrodynamic models of ejecta--CSM interaction show that $\sim$70\% of the total flux emerges in the UV at late times \citep{Dessart22}. The F275W filter samples $\sim$400~\AA\ of this UV range; $f_\mathrm{UV} = 0.3$ accounts for the strong NUV line emission (Mg~II, Fe~II) that concentrates flux near the filter pivot wavelength.} We adopt $v_\mathrm{ej} = 10{,}000$~km~s$^{-1}$, $M_\mathrm{ej} = 3~M_\odot$, and electron-scattering opacity $\kappa = 0.34$~cm$^2$~g$^{-1}$.

The model reproduces the early UV flash of SN~2023ixf \citep{Jacobson-Galan23, Hiramatsu23} for $M_\mathrm{CSM} \approx 0.01~M_\odot$ at $R_\mathrm{CSM} \approx 5 \times 10^{14}$~cm, giving $M_\mathrm{UV,peak} \approx -18.7$~mag. We note that the actual SN~2023ixf CSM is better described by a confined dense shell transitioning to a steady wind at larger radii \citep{Jacobson-Galan23}; our single-shell model captures the dominant interaction episode but does not include a wind floor. For survey-relevant parameters ($M_\mathrm{CSM} \sim 10^{-3}$--$10^{-1}~M_\odot$ at $t_\mathrm{onset} \sim 500$~d), the interaction produces $M_\mathrm{UV} \approx -13$ to $-16$ at peak and remains detectable for hundreds of days. Higher $M_\mathrm{CSM}$ produces both brighter and longer-lived emission, naturally breaking the luminosity-duration degeneracy that afflicts purely empirical parameterizations (see the SN~2023ixf model band in Figure~\ref{fig:results}).

\subsubsection{Onset time marginalization}

CSM episodes can occur at any time before explosion, placing material at arbitrary radii. The SN ejecta encounters this material at $t_\mathrm{onset} = R_\mathrm{CSM} / v_\mathrm{ej}$ days after explosion. Our survey window ($\sim$\MinPhase--\MaxPhase~days) is only sensitive to CSM at $R \sim 2 \times 10^{16}$--$2 \times 10^{17}$~cm. We cannot determine \textit{when} the CSM was ejected, only whether ejecta--CSM interaction is occurring at the time of observation. Note that $t_\mathrm{onset}$ is the post-explosion delay until the ejecta reach the shell, not the time between mass ejection and core collapse; the progenitor ejected the shell $R_\mathrm{CSM}/v_\mathrm{w}$ before explosion, typically decades to centuries for the radii probed here.

We therefore treat $t_\mathrm{onset}$ as a nuisance parameter, drawing it uniformly from $[0,\, 2000]$~days (constant episode rate) and marginalizing over it via Monte Carlo integration. For each of 500 random onset times, we evaluate the CSM UV magnitude at the observed phase, add 0.5~mag Gaussian scatter (representing viewing-angle and composition variations), and compare to the target's adopted $70\%$-coverage injection/recovery limit. The per-target detection probability $p_i$ is the fraction of MC draws that produce a detectable signal.

\subsubsection{Likelihood and priors}

The model has three free parameters:
\begin{enumerate}
\item $f_\mathrm{CSM}$: the fraction of SE~SNe with at least one CSM interaction episode (uniform prior on $[0,\,1]$).
\item $\log_{10}(M_\mathrm{CSM}/M_\odot)$: the typical CSM shell mass (uniform prior on $[-3,\,0]$, i.e.\ $10^{-3}$--$1~M_\odot$).
\item $\log_{10}(f_\mathrm{thick})$: the shell thickness fraction $\Delta R / R_\mathrm{CSM}$ (uniform prior on $[-2,\,1]$, i.e.\ $0.01$--$10$).
\end{enumerate}
All other physical parameters are fixed at the values above. The third parameter allows the data to constrain the duration of the mass-ejection episode: for a wind velocity $v_\mathrm{w}$, the ejection duration is $\Delta t_\mathrm{eject} = \Delta R / v_\mathrm{w} = f_\mathrm{thick} \times v_\mathrm{ej} \times t_\mathrm{onset} / v_\mathrm{w}$.

For non-detections, the probability of the null result is $1 - f_\mathrm{CSM} \times p_i$, where $p_i$ is the physics-based detection probability marginalized over onset time. For detections, we use the full information content of the observed magnitude: the onset-time-marginalized probability density of measuring $m_{\mathrm{obs},i}$ given the model, $\mathcal{P}_i = \langle \mathcal{N}(m_{\mathrm{obs},i} \,|\, m_{\mathrm{pred},j},\, \sigma_{\mathrm{tot},i}) \rangle_j$, where the average is over $j = 1 \ldots N_\mathrm{MC}$ onset-time draws and $\sigma_{\mathrm{tot},i} = \sqrt{\sigma_\mathrm{int}^2 + \sigma_{\mathrm{meas},i}^2}$ combines intrinsic scatter with the measurement uncertainty. The full log-likelihood is:
\begin{equation}
\ln \mathcal{L} = \sum_{i\,\in\,\mathrm{det}} \bigl[\ln f_\mathrm{CSM} + \ln \mathcal{P}_i\bigr] + \sum_{i\,\in\,\mathrm{non\text{-}det}} \ln(1 - f_\mathrm{CSM} \times p_i).
\end{equation}
This exploits the full constraining power of both detections and non-detections: the observed magnitudes of detected targets constrain $M_\mathrm{CSM}$ through the physics model, while each non-detection contributes individually through $p_i$, which encodes its specific phase, redshift, and limiting magnitude. Targets with the deepest limits (such as SN~2019yz and SN~2020zgl; Sect.~\ref{sec:nullresults}) carry the most weight, penalizing models that predict bright CSM at those phases.

Because the detection status of several targets is uncertain (Sect.~\ref{sec:wise_discussion}), we marginalize over this ambiguity by running separate MCMC chains under three detection scenarios and merging them with equal weight. A \emph{conservative} scenario uses only the five targets with the most secure CSM as detections (the three UV detections that also have WISE mid-IR excesses or ground-based interaction signatures: SN~2019oys, SN~2020jfv, and SN~2020oi, plus the two candidates SN~2019obh and SN~2020sgf with strong late-time mid-IR excess but no UV detections). The \emph{baseline} scenario uses the \Nconfirmed\ secure UV detections from Sect.~\ref{sec:classification} and treats all ambiguous candidates as non-detections. The \emph{optimistic} scenario adds the two IR-only candidates to the baseline set, yielding 5, 6, and 8 detections respectively. We use the UV limiting magnitude as the observed brightness for SN~2019obh, SN~2020sgf when these are included as detections. We ran \texttt{emcee} \citep{Foreman-Mackey13} with 32~walkers for 20{,}000~steps per scenario using differential-evolution moves \citep{terBraak06, terBraak08}, discarding the first 1{,}000~steps as burn-in. The three chains are concatenated with equal weight, yielding ${\sim}1.8$~million posterior samples with autocorrelation lengths of ${\sim}20$--30~steps.

\begin{figure*}
    \centering
    \includegraphics[width=0.95\linewidth]{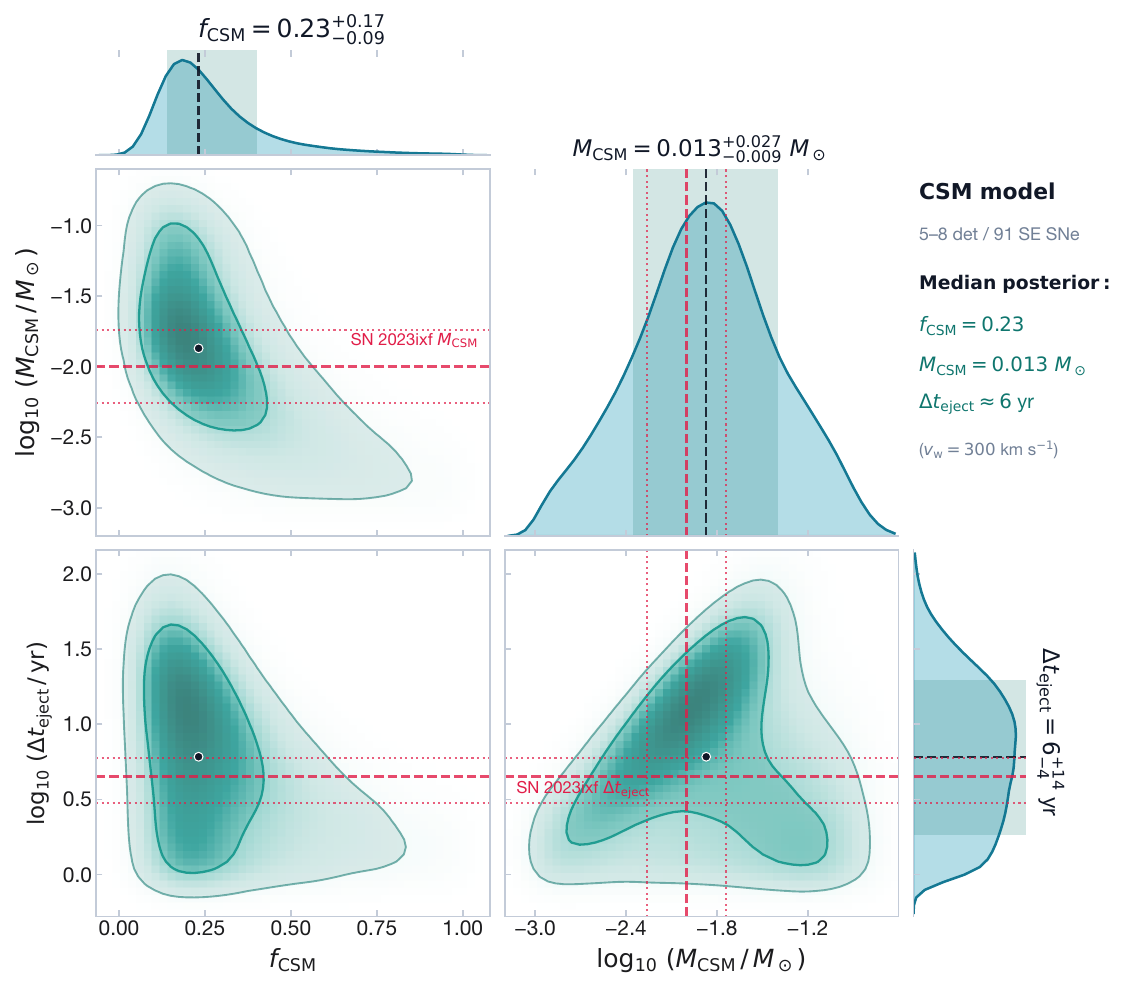}
    \caption{Corner plot of the three-parameter MCMC posterior for the CSM interaction fraction $f_\mathrm{CSM}$, shell mass $\log_{10}(M_\mathrm{CSM}/M_\odot)$, and ejection duration $\Delta t_\mathrm{eject}$ (\S\ref{sec:csmmodel}). The ejection duration is derived from the shell thickness fraction $f_\mathrm{thick}$ assuming a fiducial outflow velocity $v_\mathrm{w} = 300$~km~s$^{-1}$ (motivated by wave-driven outflow models for compact helium stars; \citealp{Fuller17, RoFuller21}) and a fiducial onset time of 1000~days; it scales inversely with $v_\mathrm{w}$. The choice of $v_\mathrm{w}$ affects only the ejection-duration axis and does not alter the posteriors for $f_\mathrm{CSM}$ or $M_\mathrm{CSM}$. Diagonal panels show 1D marginal posteriors with 68\% credible intervals (shaded). Off-diagonal panels show 2D joint posteriors with 67\% and 95\% highest-density contours. Rose dashed and dotted lines mark the median and central 67\% of the literature ranges for SN~2023ixf: $M_\mathrm{CSM} \approx 0.01~M_\odot$ \citep{Jacobson-Galan23, Hiramatsu23} and an ejection duration of $\sim$3--6~yr inferred from flash spectroscopy \citep{Jacobson-Galan23}. The SE~SN posterior ($\Delta t_\mathrm{eject} \sim \dtEjectYrThreeHundred$~yr) is broadly consistent with the SN~2023ixf timescale and rules out steady binary mass transfer ($10^3$--$10^4$~yr).}
    \label{fig:MCMC}
\end{figure*}

\subsubsection{CSM Modeling Results}

Figure~\ref{fig:MCMC} shows the joint posterior. The CSM interaction fraction is constrained to $f_\mathrm{CSM} = \fCSMmed^{+\fCSMerrplus}_{-\fCSMerrminus}$, the shell mass to $\log_{10}(M_\mathrm{CSM}/M_\odot) = \logMCSMmed^{+\logMCSMerrplus}_{-\logMCSMerrminus}$ (68\% credible intervals), corresponding to a characteristic shell mass of $M_\mathrm{CSM} \approx \MCSMsolmed~M_\odot$, and the shell thickness fraction to $\log_{10}(f_\mathrm{thick}) = \logfthickmed^{+\logfthickerrplus}_{-\logfthickerrminus}$, i.e.\ $f_\mathrm{thick} \approx \fthickmed$. The posterior on $f_\mathrm{thick}$ is $\sim$3 times narrower than its prior, demonstrating that the data meaningfully constrain the shell geometry. The preference for a geometrically thin shell ($\Delta R / R_\mathrm{CSM} \sim \fthickmed$) favors short, eruptive mass ejection over steady-state winds. The joint posterior shows a mild degeneracy: higher $M_\mathrm{CSM}$ produces brighter and longer-lived episodes, requiring a lower $f_\mathrm{CSM}$ to match the observed detection count. The shell thickness is only weakly correlated with the other two parameters. Appendix~\ref{sec:efficiency} visualizes the survey completeness as a function of $M_\mathrm{CSM}$ and onset time for the posterior median $f_\mathrm{thick}$. The inferred $f_\mathrm{CSM}$ exceeds the raw detection rate ($\ConfirmedRate\%$) by roughly a factor of three because each target is observed only once and the CSM episode may not yet have started, and because the $t^{-5/3}$ fading reduces detectability for early-onset episodes observed at late phases. Appendix~\ref{sec:simulated} shows that a simulated survey at the posterior median parameters reproduces the observed detection count and magnitude distribution.

The thin-shell posterior directly constrains the duration of the mass-ejection episode. For an outflow velocity $v_\mathrm{w}$, the ejection timescale is $\Delta t_\mathrm{eject} = f_\mathrm{thick} \times v_\mathrm{ej} \times t_\mathrm{onset} / v_\mathrm{w}$, which scales inversely with $v_\mathrm{w}$. The photometric data constrain $f_\mathrm{thick}$ but not $v_\mathrm{w}$ itself. The physical implications and the choice of fiducial $v_\mathrm{w}$ are discussed in Sect.~\ref{sec:ejection}.

%
%

\section{Discussion and Conclusions}\label{sec:conclusions}

\subsection{Ejection timescale and mass-loss mechanism}\label{sec:ejection}

The inferred thin-shell geometry ($f_\mathrm{thick} \sim \fthickmed$) implies that the CSM shells detected in our survey were ejected on timescales of years, not millennia. For a fiducial outflow velocity of $v_\mathrm{w} = 300$~km~s$^{-1}$, the posterior median $f_\mathrm{thick}$ corresponds to an ejection duration of $\sim$\dtEjectYrThreeHundred~yr, ranging from $\sim$\dtEjectYrFiveHundred~yr at $v_\mathrm{w} = 500$~km~s$^{-1}$ to $\sim$\dtEjectYrHundred~yr at $100$~km~s$^{-1}$. We adopt the fiducial $v_\mathrm{w} = 300$~km~s$^{-1}$ based on two lines of evidence. First, the wave-driven outflow models of \citet{Fuller17} and \citet{RoFuller21} predict terminal velocities of a few hundred km~s$^{-1}$ for compact helium-star progenitors, slightly above the surface escape speed. Second, the quasi-WR star HD~45166 \citep{Shenar23}, a $\sim$4~$M_\odot$ stripped helium star closely resembling SE~SN progenitors, has an observed equatorial wind velocity of $350$~km~s$^{-1}$.

Stable Roche-lobe overflow in stripped-star binaries operates on the thermal timescale of the helium star, typically $10^3$--$10^4$~yr \citep{Tauris17, Laplace20}. This is two to three orders of magnitude longer than our inferred ejection duration for any outflow velocity above $\sim$10~km~s$^{-1}$, effectively ruling out steady mass transfer as the direct origin of the CSM shells we detect. Instead, the thin-shell constraint points to an eruptive ejection mechanism. Two broad channels could operate on the required timescale, and our data do not distinguish between them:
\begin{enumerate}
\item \textit{Wave-driven mass loss.} Vigorous convection in the final nuclear burning stages can drive waves that deposit energy in the envelope and eject $\sim$$10^{-3}$--$10^{-1}~M_\odot$ on timescales of years to decades \citep{Quataert12, Fuller18}. The duration is set by the burning lifetime of the relevant shell (Ne, O, or Si), which ranges from years to decades depending on the core mass. In compact hydrogen-poor stars, however, the envelope is tightly bound and the efficiency of direct wave-driven ejection is lower than in extended Type~II progenitors \citep{Fuller18}.
\item \textit{Late-stage binary mass transfer.} Stripped helium stars can re-expand significantly during the final nuclear burning stages as the envelope adjusts to the evolving core luminosity \citep{Laplace20}. In a close binary, this re-expansion can push the progenitor back into Roche-lobe contact, initiating mass transfer that ranges from brief enhanced RLOF to dynamically unstable ejection depending on the mass ratio, orbital separation, and the rate of expansion \citep{Ivanova13}. The clock is set by the late nuclear burning stages ($\sim$years to decades) rather than by the companion's independent evolution, naturally producing the short ejection timescale we infer. This channel combines a physically motivated trigger with an efficient mass-removal mechanism that does not require wave energy alone to unbind the CSM \citep{Smith2014}.
\end{enumerate}
These two channels are not mutually exclusive: wave heating from late-stage convection \citep{Fuller18} may inflate the envelope or destabilize its outer layers, accelerating the onset of binary overflow. The present survey constrains the amount, geometry, and timing of the CSM, but not the specific launch mechanism; distinguishing between these possibilities will require radio, spectroscopy, and binary-evolution modeling targeted at the best late-time interaction cases.

The characteristic shell mass ($M_\mathrm{CSM} \approx \MCSMsolmed~M_\odot$) is comparable to the $\sim$$0.01~M_\odot$ inferred for the confined CSM of SN~2023ixf \citep{Jacobson-Galan23}, marked as a reference in Figure~\ref{fig:MCMC}. The inferred ejection duration ($\sim$\dtEjectYrThreeHundred~yr for $v_\mathrm{w} = 300$~km~s$^{-1}$) is also broadly consistent with the $\sim$3--6~yr pre-explosion mass-loss episode of SN~2023ixf inferred from flash spectroscopy \citep{Jacobson-Galan23}, though the outflow velocities differ substantially ($\sim$25~km~s$^{-1}$ for the RSG wind of SN~2023ixf vs.\ a few hundred km~s$^{-1}$ expected for compact stripped stars). The similarity in mass and timescale does not uniquely imply the same physical origin, but it is notable that both populations require eruptive mass loss in the final years before core collapse, regardless of the very different envelope structures (extended RSG vs.\ compact stripped star). We note that our constraints assume all CSM episodes produce a light curve resembling the single-shell model. The real population may be more diverse: some objects could sustain UV-bright interaction from early times, while others have genuinely no pre-explosion mass loss. Our $f_\mathrm{CSM}$ and $M_\mathrm{CSM}$ are therefore population averages conditioned on this model.

Our survey extends the emerging consensus from recent late-time HST studies \citep{BaerWay2024,Inkenhaag2025}: secure UV point-source detections remain rare, host contamination is the dominant obstacle at faint magnitudes, and the most convincing interaction cases are supported by optical, UV, IR, and spectroscopic evidence simultaneously. The UV is a particularly sensitive probe because shock-powered emission emerges strongly in Mg~II, Fe~II, and the UV continuum \citep{ChevalierFransson2017}.

\subsection{Mid-infrared detections and dust-obscured interaction}\label{sec:wise_discussion}

The mid-IR luminosity of the four SNe showing WISE excesses is not unusual for dusty interacting SNe. In our AB-magnitude system they span approximately $M_{W1}, M_{W2} \sim -17$ to $-19.5$, corresponding to $\nu L_\nu \sim \mathrm{few} \times 10^{41}$ to $\sim 10^{42}$~erg~s$^{-1}$, comparable to published late-time mid-IR measurements of known interaction-powered events such as the Type~IIn SN~2013L and SN~2017hcc \citep{Andrews2017,Chandra2022}. The four robust IR detections divide naturally into two groups. SN~2019oys and SN~2020jfv already have independent interaction evidence from optical spectroscopy, and both also have HST measurements near the WISE epoch, although the UV flux of SN~2020jfv is likely strongly contaminated by the host background. SN~2019obh and SN~2020sgf are different: both were selected because of unusual optical evolution and exhibit persistent W1/W2 excesses hundreds of days after peak, yet neither shows a secure late-time UV point source in the HST image.

This anti-correlation between mid-IR and UV emission is a natural consequence of dust. CSM interaction generates both UV photons (from the forward/reverse shocks) and warm dust emission (from pre-existing or newly formed dust heated by the shock radiation). Even a modest column of dust ($A_V \gtrsim 2$--3~mag, corresponding to $A_\mathrm{UV} \gtrsim 5$--8~mag at $\lambda \sim 2700$~\AA) is sufficient to extinguish the UV signal while leaving the 3--5~$\mu$m WISE bands largely unaffected. The implied dust masses are small ($\lesssim 10^{-3}~M_\odot$) and consistent with condensation in the post-shock cooling layer or survival of pre-existing CSM dust. Among the eight SE~SNe with secure UV or mid-IR interaction signatures, half show mid-IR emission indicative of warm dust, suggesting that dust formation or survival in CSM interaction is common but not universal among SE~SNe. That SN~2019oys and SN~2020jfv are detected in \emph{both} UV and mid-IR can be explained if dust forms in the cool dense shell between the forward and reverse shocks, interior to the UV-emitting region \citep{Sarangi2018,Takei2025}. In this geometry the UV photons escape from the shock front while newly condensed dust in the post-shock shell reradiates in the mid-IR, without requiring an aspherical CSM distribution. For these objects our dust-free CSM model overestimates the intrinsic UV luminosity, since it does not account for the fraction absorbed by dust. However, because the model ignores dust uniformly for all targets, this is a systematic bias that applies consistently across the sample and does not preferentially affect $f_\mathrm{CSM}$.

Of the 11 UV-detected SE~SNe (confirmed and ambiguous) without robust WISE excesses, 8 have late-time W2 coverage that brackets the HST observation epoch, with $3\sigma$ limits of $M_{W2} \approx -14$ to $-17$. These limits are deep enough to have detected mid-IR emission at the luminosities of the four WISE sources ($M_{W2} \approx -17.8$ to $-19.5$). The absence of WISE counterparts in these objects implies that their CSM interaction, if real, is occurring with little or no local dust. This is physically expected: radiation-hydrodynamic models show that tenuous, dust-free CSM channels the interaction power predominantly into the UV \citep{Dessart22}, and H-poor CSM (as expected for SE~SN progenitors) may be less efficient at forming or retaining dust than the H-rich shells of Type~IIn events \citep{Fox11}. A Spitzer survey of 68 SNe~IIn detected ${\sim}15\%$ at late times in the mid-IR, with the emission attributed primarily to pre-existing circumstellar dust heated by the interaction luminosity \citep{Fox11}; where no such dust reservoir exists, the interaction signature shifts entirely to UV and X-ray wavelengths. The remaining three UV-detected objects (SN~2019bgl, SN~2019tqb, and SN~2019uff) lack WISE coverage that brackets the HST observation epoch and are therefore unconstrained in the mid-IR at the time of the UV detection.

The net effect of the UV/IR dichotomy on $f_\mathrm{CSM}$ is not straightforward. On one hand, the two IR-only candidates (SN~2019obh and SN~2020sgf) suggest that dust-obscured interaction episodes are missed by the UV survey, which would push the true interaction fraction above the UV-inferred rate. On the other hand, some of the UV detections without WISE counterparts may be host-background contamination rather than genuine SN emission, which would push the fraction downward. The MCMC analysis (Sect.~\ref{sec:mcmc}) accounts for this ambiguity by marginalizing over three detection scenarios; the resulting $f_\mathrm{CSM} = \fCSMmed^{+\fCSMerrplus}_{-\fCSMerrminus}$ and its uncertainty fold in the full range of plausible UV and IR detection scenarios for our sample.

\subsection{Comparison with interaction rates at other wavelengths}\label{sec:rate_comparison}

Previous estimates of the late-time CSM interaction rate in SE~SNe come from radio and optical surveys. \citet{Margutti2017} found that ${\sim}10\%$ of Type~Ib/c SNe with late-time radio coverage (4 out of 41) showed radio rebrightening at $>500$~days, and \citet{Vinko2017} detected late-time H$\alpha$ point sources at the positions of ${\sim}13\%$ of Type~I SNe (13 out of 99, of which 10 are SE). The VLASS survey \citep{Stroh2021} found that SE~SNe dominate the late-time radio-luminous SN population (11 of 14 classified detections at 1--60~yr), but the survey is not volume-limited and does not provide a detection rate.

Our MCMC-inferred $f_\mathrm{CSM} = \fCSMmed^{+\fCSMerrplus}_{-\fCSMerrminus}$ is higher than these raw detection fractions, but this reflects two methodological differences rather than a discrepancy. First, our MCMC analysis uses a physics-based CSM model to correct for non-detection likelihoods and phase coverage, yielding an intrinsic population fraction; our raw UV-detection rate ($\Nconfirmed/\Ntotal \approx \ConfirmedRate\%$) is consistent with the radio and optical values. Second, our analysis combines UV and mid-IR data: two of our three MCMC detection scenarios include WISE-only interaction candidates that lack UV counterparts (Sect.~\ref{sec:mcmc}), boosting the effective detection count beyond what any single waveband would yield.

\subsection{Deep non-detections}\label{sec:nullresults}

The survey contains a small set of nearby SE~SNe for which the HST non-detections are physically informative in their own right. In these cases the F275W limits reach absolute magnitudes far below the secure UV counterparts, so they already rule out long-lived UV-bright interaction at the observed epoch even without reference-image subtraction. Using our CSM physics model (Sect.~\ref{sec:csmmodel}), we can translate these non-detection limits into direct upper bounds on the CSM shell mass. These constraints are fully incorporated into the MCMC analysis through the non-detection likelihood term (Eq.~1), where each target's depth and phase contribute to the joint posterior on $f_\mathrm{CSM}$ and $M_\mathrm{CSM}$.

\paragraph{SN 2019yz}

This is a Type~Ic with no HST source at 1190~days after first detection. The adopted F275W injection/recovery limit excludes any counterpart brighter than $M_{\rm F275W}=-7.0$~mag, and we find no robust W1/W2 excess. Through the CSM physics model, this limit constrains any active CSM shell to $M_\mathrm{CSM} \lesssim 2 \times 10^{-3}~M_\odot$ for onset times $\lesssim 200$~days, and more stringently ($M_\mathrm{CSM} \lesssim 5 \times 10^{-5}~M_\odot$) for onsets $\gtrsim 1000$~days. For comparison, a shell with the MCMC median mass ($M_\mathrm{CSM} = \MCSMsolmed~M_\odot$) would produce $M_\mathrm{UV} \approx -13$ at this phase, ${\sim}6$~mag brighter than the detection limit. The onset-time-marginalized detection probability at the posterior median is $59\%$, meaning that the non-detection of SN~2019yz is consistent with the median $f_\mathrm{CSM} = \fCSMmed$. The implied mass-loss rate limit is $\dot{M} \lesssim 4 \times 10^{-5}~M_\odot$~yr$^{-1}$ for $v_\mathrm{w} = 100$~km~s$^{-1}$.

\paragraph{SN 2020zgl}

This SN provides a complementary constraint at earlier phase. The HST observation was obtained 669~days after first detection and excludes any counterpart brighter than $M_{\rm F275W}=-7.0$~mag. No robust WISE excess is present. The physics model limits any active CSM shell to $M_\mathrm{CSM} \lesssim 7 \times 10^{-4}~M_\odot$ at onset times $\lesssim 200$~days; the median posterior shell mass would produce $M_\mathrm{UV} \approx -13$ at this phase. The onset-time-marginalized detection probability is ${\sim}33\%$ (lower than SN~2019yz because the earlier observation phase provides a shorter onset-time window), and the implied mass-loss rate limit is $\dot{M} \lesssim 1 \times 10^{-5}~M_\odot$~yr$^{-1}$. Together, these two nearby non-detections place the tightest individual constraints in the sample and demonstrate that at least some SE~SN progenitors undergo negligible pre-explosion mass loss, consistent with the $1 - f_\mathrm{CSM}\approx77\%$ fraction of the population inferred to lack UV-bright CSM interaction.

\subsection{Optical light curves of the detected SNe}\label{sec:lc_morphology}

Among the eight SE~SNe with secure late-time UV or mid-IR detections (six UV, two WISE-only), five have clearly unusual ZTF optical light curves: SN~2019oys exhibits a dramatic late-time rebrightening (Figure~\ref{fig:lc_spec}), SN~2020jfv remained optically active for $>2000$~days with a second peak, SN~2018hmx has an unusually broad light curve for its subtype, and the two WISE-only candidates SN~2019obh and SN~2020sgf also have anomalously long-lived optical emission. The remaining three (SN~2019tqb, SN~2020oi, and SN~2020nke) have optical light curves consistent with the canonical SE~SN population in both duration and peak brightness. Late-time UV emission in these objects would not have been predicted from their optical behavior alone, implying that CSM interaction can proceed without leaving an obvious imprint on the early-time optical light curve. This is consistent with our thin-shell model: CSM at large radii ($R \sim 10^{16}$--$10^{17}$~cm) is not reached by the ejecta until hundreds of days after explosion and therefore does not affect the early optical evolution. Among the five unusual objects, the late-time optical rebrightenings (SN~2019oys, SN~2020jfv) are also consistent with the thin-shell picture, since CSM at large radii naturally powers renewed emission at late epochs. Only the objects whose early light curves are affected (e.g., SN~2018hmx with its unusually broad profile, SN~2019obh with three peaks within the first 50~days) require an additional CSM component at smaller radii not captured by our single-shell model, or intrinsic ejecta properties unrelated to the CSM we detect at late times.

\subsection{Anomalous objects without interaction signatures}\label{sec:anomalous}

Not all unusual optical behavior in the sample is explained by CSM interaction. Among the seven SE~SNe highlighted in Figure~\ref{fig:lc_spec} for their anomalous light-curve morphology (multiple peaks, broad profiles, or slow declines), four (SN~2019oub, SN~2018dfi, SN~2019dwf, and SN~2019vsi) show neither a HST UV counterpart nor a robust late-time WISE excess. If CSM interaction were the sole driver of their unusual optical evolution, one would expect UV or mid-IR signatures at the phases probed here. The absence of both suggests that a central engine (e.g., a millisecond magnetar; \citealp{Kasen2010,Woosley2010}) or unusually massive or aspherical ejecta \citep{2023Karamehmetoglu} may contribute to the diversity of SE~SN light curves, independent of CSM. Disentangling these scenarios would have required deep UV, infrared, and radio follow-up of the optically anomalous objects that were dark in the current survey. We also note that one could have both an unusual optical LC unrelated to CSM, and a late-time CSM episode. This could be the case in e.g. SN~2019obh and SN~2018hmx (Sects.~\ref{sec:results}, and \ref{sec:lc_morphology}).

\subsection{Effect of host-galaxy extinction}\label{sec:extinction}

Our analysis corrects for Milky Way foreground extinction but not for host-galaxy reddening, which for SE~SNe is typically $E(B-V)_\mathrm{host} \lesssim 0.1$--$0.3$~mag \citep{Drout2011}, corresponding to $A_\mathrm{F275W} \approx 0.6$--$1.9$~mag. Because the apparent-magnitude detection limit is set by the image depth, a uniform host extinction shifts both detected luminosities and non-detection limits to brighter intrinsic magnitudes by the same amount, leaving $f_\mathrm{CSM}$ nearly unchanged. We verified this by re-running the full three-scenario MCMC with a uniform $E(B-V)_\mathrm{host} = 0.1$~mag ($A_\mathrm{F275W} = 0.65$~mag) applied to all targets: $f_\mathrm{CSM}$ shifted by only $+0.001$, while $\log_{10}(M_\mathrm{CSM}/M_\odot)$ increased by $+0.25$~dex to match the brighter intrinsic luminosities. Per-object host-extinction estimates from Balmer decrements or SED fitting would primarily refine $M_\mathrm{CSM}$ rather than $f_\mathrm{CSM}$.

\subsection{Summary of conclusions}\label{sec:summary}

\begin{enumerate}

\item \textbf{Late-time UV and mid-IR detections.} We detect late-time NUV emission at the SN position in \Ndetected\ of \Ntotal\ SE~SNe, of which \Nconfirmed\ (\ConfirmedRate\%) are secure point-source detections and \Nambiguous\ are ambiguous (likely host-dominated). The \Nnondet\ non-detections reach a median depth of \MedianMagLim\ AB~mag. Four SE~SNe additionally show robust late-time WISE mid-IR excesses; two of these lack UV counterparts, indicating dust-obscured interaction missed by the UV survey alone.

\item \textbf{CSM interaction fraction and shell properties.} A physics-based MCMC analysis combining detections and non-detection limits yields $f_\mathrm{CSM} = \fCSMmed^{+\fCSMerrplus}_{-\fCSMerrminus}$, with a characteristic shell mass $M_\mathrm{CSM} \approx \MCSMsolmed~M_\odot$ and thin-shell geometry ($f_\mathrm{thick} \approx \fthickmed$). The deep non-detections ($M_\mathrm{F275W} \gtrsim -7$~mag) on several nearby objects substantially tighten these constraints by ruling out bright, long-lived interaction episodes.

\item \textbf{Eruptive pre-explosion mass loss.} The thin-shell geometry implies an ejection duration of $\sim$\dtEjectYrThreeHundred~yr at the fiducial $v_\mathrm{w} = 300$~km~s$^{-1}$, two to three orders of magnitude shorter than the thermal timescale of stable Roche-lobe overflow in stripped-star binaries \citep{Tauris17}. This rules out steady mass transfer and points to eruptive ejection: wave-driven outbursts \citep{Fuller18} or binary overflow triggered by late-stage progenitor re-expansion \citep{Laplace20}, in roughly one quarter of SE~SN progenitors.

\item \textbf{Not all anomalous light curves require CSM.} Four of seven SE~SNe with the most unusual optical evolution (multiple peaks, broad profiles, or slow declines) show neither a late-time UV counterpart nor a WISE mid-IR excess, suggesting that a central engine or massive aspherical ejecta, rather than CSM interaction, can independently drive optical diversity in the SE~SN population.

\end{enumerate}

Reference-image subtraction of the existing HST data would resolve the host-contamination ambiguity for targets in the $-8$ to $-11$~mag regime, and joint UV--IR--radio modeling could further disentangle CSM interaction from central-engine power and host background. Looking ahead, the ULTRASAT wide-field NUV survey \citep{ULTRASAT2024} is expected to detect hundreds of core-collapse SNe per year in the UV, and the UVEX all-sky NUV+FUV survey \citep{UVEX2021} will reach depths $50$--$100\times$ below GALEX, together enabling population-level constraints on CSM interaction rates and timescales that go beyond the deep single-epoch limits presented here.

\acknowledgments
\small{
Based on observations obtained with the Samuel Oschin Telescope 48-inch and the 60-inch Telescope at the Palomar Observatory as part of the Zwicky Transient Facility project. ZTF is supported by the National Science Foundation under Grants No. AST-1440341, AST-2034437, and currently Award \#2407588. ZTF receives additional funding from the ZTF partnership. Current members include Caltech, USA; Caltech/IPAC, USA; University of Maryland, USA; University of California, Berkeley, USA; University of Wisconsin at Milwaukee, USA; Cornell University, USA; Drexel University, USA; University of North Carolina at Chapel Hill, USA; Institute of Science and Technology, Austria; National Central University, Taiwan, and OKC, University of Stockholm, Sweden. Operations are conducted by Caltech's Optical Observatory (COO), Caltech/IPAC, and the University of Washington at Seattle, USA.

The ZTF forced-photometry service was funded under the Heising-Simons Foundation grant \#12540303 (PI: Graham).
The Gordon and Betty Moore Foundation, through both the Data-Driven Investigator Program and a dedicated grant, provided critical funding for SkyPortal.

W.J.-G.\ is supported by NASA through Hubble Fellowship grant HST-HF2-51558.001-A awarded by the Space Telescope Science Institute, which is operated for NASA by the Association of Universities for Research in Astronomy, Inc., under contract NAS5-26555.
T.-W.C.\ acknowledges financial support from the Yushan Fellow Program of the Ministry of Education, Taiwan (MOE-111-YSFMS-0008-001-P1), and from the National Science and Technology Council, Taiwan (NSTC 114-2112-M-008-021-MY3).

Funded by the European Union (ERC, project number 101042299, TransPIre). Views and opinions expressed are however those of the author(s) only and do not necessarily reflect those of the European Union or the European Research Council Executive Agency. Neither the European Union nor the granting authority can be held responsible for them.








Partially based on observations made with the Nordic Optical Telescope, operated by the Nordic Optical Telescope Scientific Association at the Observatorio del Roque de los Muchachos, La Palma, Spain, of the Instituto de Astrofisica de Canarias. Some of the data presented here were obtained with ALFOSC, which is provided by the Instituto de Astrofisica de Andalucia (IAA) under a joint agreement with the University of Copenhagen and NOTSA.

Some of the data presented herein were obtained at the W. M. Keck Observatory, which is operated as a scientific partnership among the California Institute of Technology, the University of California, and NASA; the observatory was made possible by the generous financial support of the W. M. Keck Foundation. 


The Liverpool Telescope is operated on the island of La Palma by Liverpool John Moores University in the Spanish Observatorio del Roque de los Muchachos of the Instituto de Astrofisica de Canarias with financial support from the UK Science and Technology Facilities Council.

The SED Machine is based upon work supported by the National Science Foundation under Grant No. 1106171. 
}

\software{\texttt{astropy} \citep{Astropy-Collaboration13},
          \texttt{numpy} \citep{Harris20},
          \texttt{scipy} \citep{Jones01},
          \texttt{matplotlib} \citep{Hunter07},
          \texttt{pandas} \citep{McKinney10},
          \texttt{emcee} \citep{Foreman-Mackey13},
          \texttt{corner} \citep{Foreman-Mackey16},
          \texttt{photutils} \citep{Bradley23},
          \texttt{astroquery} \citep{Ginsburg19},
          \texttt{astroscrappy} \citep{vanDokkum01},
          \texttt{WebbPSF} \citep{Perrin2014},
          \texttt{SExtractor} \citep{sextractor}.
          }

This publication makes use of data products from the Near-Earth Object Wide-field Infrared Survey Explorer (NEOWISE), which is a joint project of the Jet Propulsion Laboratory/California Institute of Technology and the University of California, Los Angeles. NEOWISE is funded by the National Aeronautics and Space Administration.

\appendix

\section{Notes on Individual UV Candidates}\label{sec:candidate_notes}
This appendix gives a brief image-based note for every secure or ambiguous HST candidate counterpart. The adopted primary counterpart is the FLC-confirmed source with the smallest ellipse-normalized astrometric offset. Ambiguous cases either contain a second comparably likely in-ellipse source or fail the point-source morphology check. Table~\ref{tab:candidate_logic} summarizes the formal decision terms for each retained candidate, including the two-frame FLC recovery, the close-rival check, the PSF morphology flag, and the host-confusion metrics.

\startlongtable
\begin{deluxetable*}{llcccccccccc}
\tabletypesize{\scriptsize}
\tablecaption{Decision Criteria for the Retained UV Candidates\label{tab:candidate_logic}}
\tablewidth{0pt}
\tablehead{
\colhead{ZTF name} & \colhead{IAU name} & \colhead{Status} & \colhead{$N_{\rm raw}/N_{\rm FLC}$} & \colhead{FLC1} & \colhead{FLC2} & \colhead{Rival} & \colhead{PSF} & \colhead{$p(\mathrm{HII})$} & \colhead{$N_{\rm exp}$} & \colhead{Host} & \colhead{Trigger}
}
\startdata
  ZTF18accnoli & SN 2018hmx & detection & 1/1 & 3.9 & 4.9 & N & 0.86 & 0.38 & 0.08 & N & clean \\
  ZTF19aakpcuw & SN 2019bgl & ambiguous & 1/1 & 8.3 & 8.6 & N & 0.74 & 0.50 & 0.23 & N & p(HII) \\
  ZTF19abucwzt & SN 2019oys & detection & 1/1 & 124.0 & 136.7 & N & 1.00 & 0.18 & 0.14 & N & clean \\
  ZTF19acjtpqd & SN 2019tqb & detection & 2/1 & 155.8 & 159.1 & N & 1.00 & 0.08 & 0.19 & N & clean \\
  ZTF19acmelor & SN 2019uff & ambiguous & 2/1 & 64.9 & 7.3 & N & 0.85 & 0.53 & 0.13 & N & p(HII) \\
  ZTF20aaelulu & SN 2020oi & detection & 2/2 & 222.4 & 212.0 & N & 1.00 & 0.19 & 0.29 & N & clean \\
  ZTF20aahbamv & SN 2020amv & detection & 2/1 & 22.0 & 26.9 & N & 0.99 & 0.17 & 0.19 & N & clean \\
  ZTF20abbplei & SN 2020lao & ambiguous & 1/1 & 3.6 & 5.6 & N & 0.65* & 0.55 & 0.18 & N & psf+p(HII) \\
  ZTF20abgbuly & SN 2020jfv & detection & 1/1 & 439.2 & 538.0 & N & 1.00 & 0.05 & 0.13 & N & clean \\
  ZTF20abhlncz & SN 2020nke & detection & 1/1 & 8.9 & 12.4 & N & 0.91 & 0.38 & 0.15 & N & clean \\
  ZTF20abodalh & SN 2020qbu & ambiguous & 1/1 & 6.8 & 10.6 & N & 0.66* & 0.73 & 0.23 & N & psf+p(HII) \\
  ZTF20acjpvyd & SN 2020tjd & ambiguous & 1/1 & 3.8 & 4.6 & N & 0.80 & 0.91 & 0.13 & N & p(HII) \\
  ZTF20adadrhw & SN 2020adow & ambiguous & 1/1 & 11.9 & 135.2 & N & 0.90 & 0.86 & 0.67 & Y & p(HII)+host \\
  ZTF21aaqvsvw & SN 2021heh & ambiguous & 1/1 & 14.6 & 15.4 & N & 0.63* & 0.41 & 0.19 & N & psf \\
\enddata
\tablecomments{The table lists every retained HST UV candidate (confirmed or ambiguous) and the decision terms used by the classifier. $N_{\rm raw}/N_{\rm FLC}$ gives the number of raw DRC detections inside the astrometric ellipse and the number of those sources that are recovered at $\geq 3\sigma$ in both individual FLC exposures. FLC1 and FLC2 are the per-frame S/N values for the adopted primary candidate. Rival = Y means that a second FLC-confirmed source inside the ellipse has a positional likelihood at least half that of the primary. The PSF column gives the primary-candidate PSF-fit $R^2$; an asterisk marks the morphology flag used to downgrade visibly non-pointlike or blended sources. $N_{\rm exp}$ is the Poisson expectation value of unrelated compact control sources inside the astrometric ellipse, computed from the local control-source surface density. Host = Y corresponds to the formal host-confusion downgrade branch, which requires both $p(\mathrm{HII})>0.5$ and $N_{\rm exp}>0.5$. Trigger summarizes which classification branch controls the final status: ``rival'' for competing positional matches, ``psf'' for morphology flags, ``p(HII)'' when the candidate luminosity is consistent with an HII region ($p(\mathrm{HII})>0.5$), and ``host'' when the full host-confusion criterion is met.}
\end{deluxetable*}

\paragraph{SN~2018hmx (ZTF18accnoli).} A single FLC-confirmed source is present in the HST image, and no second in-ellipse candidate competes with it; the association is therefore secure, although the local background is structured.
\paragraph{SN~2019oys (ZTF19abucwzt).} A single compact FLC-confirmed source lies close to the expected SN position. Together with the late-time optical, radio, and WISE evidence discussed in the main text, this is one of the clearest interaction cases in the sample.
\paragraph{SN~2019tqb (ZTF19acjtpqd).} The HST image contains two FLC-confirmed sources inside the astrometric ellipse, but the secondary source has a PSF goodness-of-fit $R^{2} = 0.17$, well below the $R^{2}\geq0.60$ threshold required for a viable rival (Sect.~\ref{sec:classification}). After filtering, only the primary candidate remains ($R^{2} = 1.00$), and it is adopted as a secure counterpart.
\paragraph{SN~2020oi (ZTF20aaelulu).} Two FLC-confirmed sources are present inside the ellipse, but one is distinctly favored by the astrometric ranking and is adopted as the primary counterpart. The source is therefore secure as an HST detection, although the crowded nuclear environment still complicates the interpretation.
\paragraph{SN~2020amv (ZTF20aahbamv).} There is a single FLC-confirmed source inside the current astrometry ellipse. This detection is expected since this is a known interacting Type~II comparison object. The late-time UV source is therefore retained as secure.
\paragraph{SN~2020jfv (ZTF20abgbuly).} A bright unique FLC-confirmed source is present at the SN position. Together with the strong late-time optical and WISE evidence, this is a secure interaction detection.
\paragraph{SN~2020nke (ZTF20abhlncz).} A single FLC-confirmed in-ellipse source with no competing counterpart and good PSF morphology ($R^2 = 0.91$). The source lies in the faint regime where host contamination remains possible, but the association is secure.

\paragraph{SN~2019bgl (ZTF19aakpcuw).} Only one FLC-confirmed in-ellipse source is present, but its morphology is not cleanly point-like in the PSF diagnostics and $p(\mathrm{HII}) = 0.50$, so we classify the object as ambiguous.
\paragraph{SN~2019uff (ZTF19acmelor).} The HST image contains one FLC-confirmed in-ellipse source near the edge of the allowed region. It is unique in the astrometric sense, but $p(\mathrm{HII}) = 0.53$ places it just above the host-confusion threshold, so we retain it as ambiguous.
\paragraph{SN~2020lao (ZTF20abbplei).} A single faint FLC-confirmed source is present, but its PSF morphology is not cleanly point-like ($R^2 = 0.65$) and the source appears blended with the local background ($p(\mathrm{HII}) = 0.55$). We retain it as ambiguous.
\paragraph{SN~2020qbu (ZTF20abodalh).} A single FLC-confirmed source is present, and no second source competes with it astrometrically. However, the source appears morphologically blended and sits on a bright host background ($p(\mathrm{HII}) = 0.73$), so the physical interpretation is less secure than for the strongest detections.
\paragraph{SN~2020tjd (ZTF20acjpvyd).} The HST image contains one FLC-confirmed in-ellipse source. Its $p(\mathrm{HII}) = 0.91$ indicates the source is very likely an underlying host region rather than a SN counterpart; we retain it as ambiguous.
\paragraph{SN~2020adow (ZTF20adadrhw).} A single faint FLC-confirmed source ($M_{\rm UV}\approx-8$) is present near the edge of the ellipse. The source is unique within the search region, but $p(\mathrm{HII}) = 0.86$ and the faint absolute magnitude, well below the other secure counterparts, indicate probable host background rather than a SN counterpart, making this an ambiguous case.
\paragraph{SN~2021heh (ZTF21aaqvsvw).} A single FLC-confirmed source is present near the edge of the astrometric ellipse, but its PSF morphology is not cleanly point-like ($R^2 = 0.63$) and the source appears blended with the local host structure. Despite a moderate $p(\mathrm{HII}) = 0.41$, the morphology flag places it in the ambiguous category.

\section{Photometry Tables}\label{sec:phottable}

\startlongtable
\begin{deluxetable*}{llcccccccccl}
\tablecaption{HST UV Photometry of the SE~SN Sample and the Retained Type~II Comparison Object\label{tab:targets}}
\tablehead{
\colhead{ZTF name} & \colhead{IAU name} & \colhead{Type} & \colhead{$z$} & \colhead{Filter} & \colhead{$E(B\!-\!V)$} & \colhead{$m_{\rm PSF}$} & \colhead{$m_{\rm lim}$} & \colhead{$M_{\rm UV}$} & \colhead{Phase} & \colhead{Status} & \colhead{Ref.}
}
\startdata
  ZTF18aalrxas & SN 2018mpl & IIb & $0.0588$ & F275W & $0.019$ & --- & $25.73$ & --- & $1559$ & $<$ & 1 \\
  ZTF18aarpqof & SN 2018cfh & Ic & $0.0294$ & F275W & $0.015$ & --- & $25.58$ & --- & $1423$ & $<$ &  \\
  ZTF18aasycpd & SN 2018bho & IIb & $0.0450$ & F275W & $0.023$ & --- & $26.12$ & --- & $1379$ & $<$ &  \\
  ZTF18aaxiuyp & SN 2018cem & Ic & $0.0303$ & F275W & $0.030$ & --- & $25.11$ & --- & $1423$ & $<$ &  \\
  ZTF18abbovuc & SN 2018cve & Ic & $0.0390$ & F275W & $0.020$ & --- & $26.07$ & --- & $1466$ & $<$ &  \\
  ZTF18abdkkwa & SN 2018dgx & Ic & $0.0253$ & F275W & $0.036$ & --- & $26.25$ & --- & $1366$ & $<$ &  \\
  ZTF18abesqnb & SN 2018dyh & IIb & $0.0700$ & F275W & $0.018$ & --- & $25.88$ & --- & $1845$ & $<$ &  \\
  ZTF18abffyqp & SN 2018dfi & IIb & $0.0313$ & F275W & $0.019$ & --- & $26.43$ & --- & $1552$ & $<$ &  \\
  ZTF18abfzfcv & SN 2018dzw & IIb & $0.0380$ & F275W & $0.011$ & --- & $25.88$ & --- & $1368$ & $<$ &  \\
  ZTF18abjrbza & SN 2018ebt & Ic & $0.0095$ & F275W & $0.247$ & --- & $25.73$ & --- & $1329$ & $<$ &  \\
  ZTF18abktmfz & SN 2018eoe & Ib & $0.0205$ & F275W & $0.054$ & --- & $25.93$ & --- & $1480$ & $<$ &  \\
  ZTF18abwkrbl & SN 2018gjx & IIb & $0.0120$ & F275W & $0.076$ & --- & $25.58$ & --- & $1818$ & $<$ & 2 \\
  ZTF18acbzvpg & SN 2018iby & IIb & $0.0260$ & F275W & $0.019$ & --- & $25.86$ & --- & $1208$ & $<$ &  \\
  ZTF18accnoli & SN 2018hmx & IIb & $0.0380$ & F275W & $0.042$ & $25.68$ & $26.05$ & $-10.8$ & $1213$ & $\bigstar$ &  \\
  ZTF18acqxyiq & SN 2018jak & IIb & $0.0385$ & F275W & $0.011$ & --- & $25.96$ & --- & $1415$ & $<$ &  \\
  ZTF19aadnhdo & SN 2019us & Ic & $0.0298$ & F275W & $0.049$ & --- & $25.60$ & --- & $1327$ & $<$ &  \\
  ZTF19aadttht & SN 2019yz & Ic & $0.0064$ & F275W & $0.098$ & --- & $25.86$ & --- & $1190$ & $<$ &  \\
  ZTF19aakirwj & SN 2019amm & Ib & $0.0316$ & F275W & $0.035$ & --- & $25.91$ & --- & $1252$ & $<$ &  \\
  ZTF19aaknate & SN 2019bao & IIb & $0.0119$ & F275W & $0.025$ & --- & $26.02$ & --- & $1221$ & $<$ &  \\
  ZTF19aakpcuw & SN 2019bgl & Ic & $0.0306$ & F275W & $0.024$ & --- & $25.79$ & --- & $1323$ & $\sim$ &  \\
  ZTF19aalouag & SN 2019buy & Ib & $0.0549$ & F275W & $0.012$ & --- & $25.95$ & --- & $1320$ & $<$ &  \\
  ZTF19aamgghn & SN 2019bwi & Ib & $0.0297$ & F275W & $0.091$ & --- & $25.82$ & --- & $1127$ & $<$ &  \\
  ZTF19aamsetj & SN 2019cad & Ic & $0.0275$ & F275W & $0.015$ & --- & $26.10$ & --- & $1055$ & $<$ & 4 \\
  ZTF19aarfkch & SN 2019dwf & IIb & $0.0510$ & F275W & $0.019$ & --- & $26.24$ & --- & $959$ & $<$ &  \\
  ZTF19aatheus & SN 2019eix & Ic & $0.0184$ & F275W & $0.060$ & --- & $25.80$ & --- & $998$ & $<$ & 3 \\
  ZTF19aaxfcpq & SN 2019gwc & Ic-BL & $0.0380$ & F275W & $0.011$ & --- & $26.27$ & --- & $1556$ & $<$ &  \\
  ZTF19abacxod & SN 2019hvg & IIb & $0.0176$ & F275W & $0.023$ & --- & $25.92$ & --- & $1014$ & $<$ &  \\
  ZTF19abafmwj & SN 2019icf & Ib & $0.0340$ & F275W & $0.028$ & --- & $25.96$ & --- & $1070$ & $<$ &  \\
  ZTF19abbthwy & SN 2019igh & IIb & $0.0250$ & F275W & $0.023$ & --- & $26.08$ & --- & $1019$ & $<$ &  \\
  ZTF19abdoior & SN 2019krw & Ic & $0.0470$ & F275W & $0.038$ & --- & $26.01$ & --- & $948$ & $<$ &  \\
  ZTF19abfsxpw & SN 2019lci & Ic-BL & $0.0290$ & F275W & $0.066$ & --- & $25.80$ & --- & $1017$ & $<$ &  \\
  ZTF19abgfuhh & SN 2019lgc & IIb-pec & $0.0354$ & F275W & $0.062$ & --- & $25.87$ & --- & $1115$ & $<$ &  \\
  ZTF19abgqruu & AT 2019mjo & Ib & $0.0407$ & F275W & $0.018$ & --- & $25.77$ & --- & $1165$ & $<$ &  \\
  ZTF19abqmsnk & SN 2019ofk & IIb & $0.0356$ & F275W & $0.040$ & --- & $25.86$ & --- & $1074$ & $<$ &  \\
  ZTF19abqwtfu & SN 2019odp & Ib & $0.0144$ & F275W & $0.167$ & --- & $25.68$ & --- & $1103$ & $<$ & 5 \\
  ZTF19abqykei & SN 2019obh & IIb & $0.0355$ & F275W & $0.065$ & --- & $25.78$ & --- & $1145$ & $<$ &  \\
  ZTF19abtsnyy & SN 2019oub & Ic & $0.0400$ & F275W & $0.308$ & --- & $26.18$ & --- & $952$ & $<$ &  \\
  ZTF19abucwzt & SN 2019oys & Ib & $0.0162$ & F275W & $0.078$ & $21.94$ & $26.10$ & $-12.9$ & $1127$ & $\bigstar$ & 6,7 \\
  ZTF19abxjrge & SN 2019ply & IIb & $0.0217$ & F275W & $0.126$ & --- & $26.08$ & --- & $1464$ & $<$ &  \\
  ZTF19abxphdh & AT 2019pnm & Ic & $0.0355$ & F275W & $0.074$ & --- & $25.74$ & --- & $1084$ & $<$ &  \\
  ZTF19abxtcio & SN 2019pof & Ib & $0.0155$ & F275W & $0.052$ & --- & $25.89$ & --- & $1459$ & $<$ &  \\
  ZTF19abzwaen & SN 2019qfi & Ic-BL & $0.0285$ & F275W & $0.060$ & --- & $25.93$ & --- & $1094$ & $<$ &  \\
  ZTF19acbvbtz & AT 2019rpt & Ib & $0.0180$ & F275W & $0.040$ & --- & $26.08$ & --- & $1096$ & $<$ &  \\
  ZTF19acfejbj & SN 2019sox & IIb & $0.0109$ & F275W & $0.049$ & --- & $25.88$ & --- & $916$ & $<$ &  \\
  ZTF19acjtpqd & SN 2019tqb & Ic & $0.0172$ & F275W & $0.007$ & $20.92$ & $26.14$ & $-13.6$ & $1110$ & $\bigstar$ &  \\
  ZTF19ackjene & SN 2019tzx & IIb & $0.0435$ & F275W & $0.075$ & --- & $25.96$ & --- & $901$ & $<$ &  \\
  ZTF19ackjjwf & SN 2019tls & Ib & $0.0162$ & F275W & $0.027$ & --- & $26.18$ & --- & $890$ & $<$ &  \\
  ZTF19acmelor & SN 2019uff & Ib & $0.0266$ & F275W & $0.029$ & --- & $26.12$ & --- & $1030$ & $\sim$ &  \\
  ZTF19acxpuql & SN 2019vsi & Ib & $0.0280$ & F275W & $0.013$ & --- & $25.67$ & --- & $833$ & $<$ &  \\
  ZTF19acyogrm & SN 2019wzj & Ic & $0.0200$ & F275W & $0.161$ & --- & $25.88$ & --- & $988$ & $<$ &  \\
  ZTF19adbrjic & SN 2019xuk & Ib & $0.0350$ & F275W & $0.050$ & --- & $25.90$ & --- & $812$ & $<$ &  \\
  ZTF20aaaxghv & SN 2019yfv & Ib & $0.0400$ & F275W & $0.032$ & --- & $25.71$ & --- & $762$ & $<$ &  \\
  ZTF20aaelulu & SN 2020oi & Ic & $0.0052$ & F275W & $0.022$ & $19.56$ & $25.75$ & $-12.4$ & $803$ & $\bigstar$ & 8,9,10,11 \\
  ZTF20aaerxne & SN 2019yxp & Ib & $0.0104$ & F275W & $0.011$ & --- & $25.94$ & --- & $817$ & $<$ &  \\
  ZTF20aafmdzj & SN 2020zg & Ic-BL & $0.0557$ & F275W & $0.027$ & --- & $25.89$ & --- & $951$ & $<$ &  \\
  ZTF20aahbamv & SN 2020amv & II & $0.0452$ & F275W & $0.031$ & $24.02$ & $26.01$ & $-12.8$ & $797$ & $\bigstar$ & 12 \\
  ZTF20aahdutc & SN 2020akf & Ic & $0.0124$ & F275W & $0.013$ & --- & $25.99$ & --- & $676$ & $<$ &  \\
  ZTF20aahfqpm & SN 2020ano & IIb & $0.0311$ & F275W & $0.018$ & --- & $26.15$ & --- & $753$ & $<$ &  \\
  ZTF20aaiqiti & SN 2020ayz & Ic-BL & $0.0250$ & F275W & $0.012$ & --- & $25.90$ & --- & $1152$ & $<$ &  \\
  ZTF20aalcyih & SN 2020bpf & Ib & $0.0175$ & F275W & $0.064$ & --- & $25.98$ & --- & $689$ & $<$ &  \\
  ZTF20aammtwx & SN 2020cgu & Ic & $0.0272$ & F275W & $0.049$ & --- & $26.04$ & --- & $658$ & $<$ &  \\
  ZTF20aanvmdt & SN 2020cpg & Ib & $0.0369$ & F275W & $0.026$ & --- & $25.87$ & --- & $795$ & $<$ & 13 \\
  ZTF20aavgmli & SN 2020htt & Ib & $0.0230$ & F275W & $0.056$ & --- & $25.88$ & --- & $590$ & $<$ &  \\
  ZTF20abaszgh & SN 2020ksa & Ib & $0.0222$ & F275W & $0.016$ & --- & $25.52$ & --- & $613$ & $<$ &  \\
  ZTF20abbplei & SN 2020lao & Ic-BL & $0.0312$ & F275W & $0.044$ & --- & $25.89$ & --- & $627$ & $\sim$ &  \\
  ZTF20abdajuq & AT 2020lma & Ic & $0.0411$ & F275W & $0.060$ & --- & $25.63$ & --- & $585$ & $<$ &  \\
  ZTF20abevbxv & SN 2020mmz & IIb & $0.0053$ & F275W & $0.047$ & --- & $25.90$ & --- & $841$ & $<$ &  \\
  ZTF20abgbuly & SN 2020jfv & IIb & $0.0171$ & F275W & $0.044$ & $20.27$ & $26.93$ & $-14.4$ & $839$ & $\bigstar$ & 12 \\
  ZTF20abgbxvb & SN 2020nac & Ib & $0.0174$ & F275W & $0.211$ & --- & $24.76$ & --- & $526$ & $<$ &  \\
  ZTF20abhlncz & SN 2020nke & Ib & $0.0308$ & F275W & $0.054$ & $25.19$ & $26.64$ & $-10.9$ & $765$ & $\bigstar$ &  \\
  ZTF20abkiarz & SN 2020oce & Ic & $0.0235$ & F275W & $0.067$ & --- & $25.88$ & --- & $900$ & $<$ &  \\
  ZTF20abodalh & SN 2020qbu & Ic & $0.0200$ & F275W & $0.018$ & --- & $25.88$ & --- & $489$ & $\sim$ &  \\
  ZTF20abwzqzo & SN 2020sbw & IIb & $0.0230$ & F275W & $0.033$ & --- & $25.98$ & --- & $770$ & $<$ &  \\
  ZTF20abxpoxd & SN 2020sgf & Ib & $0.0221$ & F275W & $0.182$ & --- & $25.79$ & --- & $720$ & $<$ &  \\
  ZTF20abxyaaf & SN 2020rur & Ic & $0.0156$ & F275W & $0.076$ & --- & $25.73$ & --- & $721$ & $<$ &  \\
  ZTF20abywsut & SN 2020sya & Ic & $0.0314$ & F275W & $0.051$ & --- & $25.86$ & --- & $747$ & $<$ &  \\
  ZTF20acjpvyd & SN 2020tjd & IIb & $0.0100$ & F275W & $0.029$ & --- & $26.11$ & --- & $690$ & $\sim$ &  \\
  ZTF20aclvvws & SN 2020ybq & Ib & $0.0255$ & F275W & $0.079$ & --- & $26.25$ & --- & $674$ & $<$ &  \\
  ZTF20acpjqxp & SN 2020zgl & Ib & $0.0065$ & F275W & $0.050$ & --- & $25.60$ & --- & $669$ & $<$ &  \\
  ZTF20actqnhg & SN 2020aaxf & IIb & $0.0148$ & F275W & $0.067$ & --- & $24.99$ & --- & $669$ & $<$ &  \\
  ZTF20acwobku & SN 2020acct & IIb-pec & $0.0347$ & F275W & $0.026$ & --- & $25.81$ & --- & $354$ & $<$ & 14 \\
  ZTF20acwofhd & SN 2020acfp & Ic & $0.0328$ & F275W & $0.018$ & --- & $25.19$ & --- & $664$ & $<$ &  \\
  ZTF20acwqqjs & SN 2020acat & IIb & $0.0079$ & F275W & $0.020$ & --- & $26.06$ & --- & $464$ & $<$ & 15,16 \\
  ZTF20adadrhw & SN 2020adow & Ic-BL & $0.0075$ & F275W & $0.031$ & --- & $26.55$ & --- & $395$ & $\sim$ &  \\
  ZTF21aaabrpu & SN 2021rf & IIb & $0.0272$ & F275W & $0.012$ & --- & $25.93$ & --- & $394$ & $<$ &  \\
  ZTF21aaaosoq & SN 2021bm & Ic & $0.0315$ & F275W & $0.007$ & --- & $25.91$ & --- & $473$ & $<$ &  \\
  ZTF21aaaubig & SN 2021do & Ic & $0.0093$ & F275W & $0.021$ & --- & $25.91$ & --- & $569$ & $<$ &  \\
  ZTF21aacufip & SN 2021vz & Ic & $0.0450$ & F275W & $0.016$ & --- & $25.98$ & --- & $585$ & $<$ &  \\
  ZTF21aaoiaar & SN 2021esi & IIb & $0.0330$ & F275W & $0.078$ & --- & $25.92$ & --- & $270$ & $<$ &  \\
  ZTF21aapfnoc & AT 2021fyq & Ib & $0.0190$ & F275W & $0.013$ & --- & $25.89$ & --- & $578$ & $<$ &  \\
  ZTF21aaqvsvw & SN 2021heh & IIb & $0.0266$ & F275W & $0.031$ & --- & $26.65$ & --- & $369$ & $\sim$ &  \\
  ZTF21aarnjyd & SN 2021hrj & Ib & $0.0220$ & F275W & $0.093$ & --- & $25.92$ & --- & $320$ & $<$ &  \\
\enddata
\tablerefs{(1)~\citet{Fremling_aalrxas}; (2)~\citet{Prentice2020_2018gjx}; (3)~\citet{PadillaGonzalez2023}; (4)~\citet{Gutierrez2021}; (5)~\citet{Schweyer2025}; (6)~\citet{Sollerman_2020}; (7)~\citet{Sfaradi_2024}; (8)~\citet{Gagliano_2022}; (9)~\citet{Horesh2020}; (10)~\citet{Maeda2021}; (11)~\citet{Rho2021}; (12)~\citet{Sollerman_2021}; (13)~\citet{Medler2021}; (14)~\citet{Angus2024}; (15)~\citet{Medler2022}; (16)~\citet{Ergon2024}.}
\tablecomments{Phase is measured in days from the first $5\sigma$ detection in ZTF forced photometry. $E(B\!-\!V)$ is the Milky Way foreground reddening from \citet{Schlafly11}. The apparent magnitude $m_{\rm PSF}$ is the observed (not extinction-corrected) PSF-fit AB magnitude. The absolute magnitude $M_{\rm UV}$ is corrected for Milky Way extinction using $A_\lambda = R_\lambda \times E(B\!-\!V)$ with $R_{\rm F275W} = 6.47$ and $R_{\rm F336W} = 5.04$ from \citet{Fitzpatrick99} and \citet{Schlafly11}. Confirmed detections ($\bigstar$) report the PSF-fit magnitude of the adopted primary candidate, selected as the FLC-confirmed source with the smallest astrometric ellipse-normalized offset. Ambiguous cases ($\sim$) have more than one comparably plausible in-ellipse counterpart or a morphologically non-pointlike primary, so no unique UV magnitude is reported. The limit column reports the adopted $70\%$-coverage astrometry-marginalized PSF injection/recovery limit used in the main analysis.}
\end{deluxetable*}

\begin{deluxetable*}{llccccccc}
\tablecaption{Robust late-time WISE detections in the HST SE~SN sample\label{tab:wise}}
\tablehead{
\colhead{ZTF name} & \colhead{IAU name} & \colhead{Type} & \colhead{Phase range} & \colhead{$\langle t \rangle$} & \colhead{$\langle M_{W1} \rangle$} & \colhead{$\langle M_{W2} \rangle$} & \colhead{HST UV det.} & \colhead{Note}
}
\startdata
  ZTF20abgbuly & SN 2020jfv & IIb & $356$--$883$ & $625$ & $-17.92$ & $-18.27$ & yes & known interaction \\
  ZTF19abucwzt & SN 2019oys & Ib & $407$--$1137$ & $760$ & $-17.21$ & $-17.76$ & yes & known interaction \\
  ZTF20abxpoxd & SN 2020sgf & Ib & $362$--$726$ & $533$ & --- & $-17.79$ & no & IR-selected candidate \\
  ZTF19abqykei & SN 2019obh & IIb & $376$--$1107$ & $732$ & $-18.51$ & $-19.48$ & no & IR-selected candidate \\
\enddata
\tablecomments{Phases are rest-frame days relative to the optical peak. $\langle t \rangle$ is the inverse-variance-weighted mean late-time WISE epoch, and $\langle M_{W1} \rangle$ and $\langle M_{W2} \rangle$ are the corresponding baseline-subtracted absolute AB magnitudes of the late-time excess. ``HST UV det.'' indicates whether the object has a secure late-time HST UV point-source detection at the SN position, not merely whether it was observed by HST.}
\end{deluxetable*}


\startlongtable
\begin{deluxetable*}{llcccccccc}
\tablecaption{Astrometric Uncertainty Budget\label{tab:astrometry}}
\tablewidth{0pt}
\tablehead{
\colhead{ZTF name} & \colhead{IAU name} & \colhead{$N_{\rm alerts}$} & \colhead{$\sigma_{\rm ZTF,sys}$} & \colhead{$\sigma_{\rm ZTF,stat}$} & \colhead{$\sigma_{\rm HST}$} & \colhead{$N_{\rm Gaia}$} & \colhead{$\sigma_{\rm tot}$} & \colhead{$3.5\sigma$ search} \\
\colhead{} & \colhead{} & \colhead{} & \colhead{(RA, DEC)} & \colhead{(RA, DEC)} & \colhead{(RA, DEC)} & \colhead{} & \colhead{(RA, DEC)} & \colhead{(RA$\times$DEC)}
}
\startdata
  ZTF18aalrxas & SN 2018mpl & 260 & 0.06, 0.06 & 0.011, 0.019 & 0.116, 0.030 & 3 & 0.131, 0.070 & $0.46''\times 0.24''$ \\
  ZTF18aarpqof & SN 2018cfh & 235 & 0.06, 0.06 & 0.019, 0.022 & 0.123, 0.046 & 4 & 0.138, 0.079 & $0.48''\times 0.28''$ \\
  ZTF18aasycpd & SN 2018bho & 37 & 0.06, 0.06 & 0.008, 0.017 & 0.111, 0.049 & 3 & 0.126, 0.079 & $0.44''\times 0.28''$ \\
  ZTF18aaxiuyp & SN 2018cem & 253 & 0.06, 0.06 & 0.006, 0.008 & 0.038, 0.096 & 7 & 0.071, 0.114 & $0.25''\times 0.40''$ \\
  ZTF18abbovuc & SN 2018cve & 27 & 0.06, 0.06 & 0.019, 0.019 & 0.078, 0.030 & 2 & 0.100, 0.070 & $0.35''\times 0.24''$ \\
  ZTF18abdkkwa & SN 2018dgx & 312 & 0.06, 0.06 & 0.005, 0.004 & 0.021, 0.018 & 5 & 0.064, 0.063 & $0.22''\times 0.22''$ \\
  ZTF18abesqnb & SN 2018dyh & 112 & 0.06, 0.06 & 0.006, 0.007 & 0.089, 0.121 & 2 & 0.108, 0.135 & $0.38''\times 0.47''$ \\
  ZTF18abffyqp & SN 2018dfi & 313 & 0.06, 0.06 & 0.005, 0.005 & 0.019, 0.034 & 5 & 0.063, 0.069 & $0.22''\times 0.24''$ \\
  ZTF18abfzfcv & SN 2018dzw & 286 & 0.06, 0.06 & 0.008, 0.011 & 0.057, 0.065 & 6 & 0.083, 0.089 & $0.29''\times 0.31''$ \\
  ZTF18abjrbza & SN 2018ebt & 110 & 0.06, 0.06 & 0.002, 0.007 & 0.076, 0.068 & 11 & 0.097, 0.091 & $0.34''\times 0.32''$ \\
  ZTF18abktmfz & SN 2018eoe & 383 & 0.06, 0.06 & 0.004, 0.005 & 0.062, 0.038 & 7 & 0.086, 0.071 & $0.30''\times 0.25''$ \\
  ZTF18abwkrbl & SN 2018gjx & 56 & 0.06, 0.06 & 0.004, 0.003 & 0.030, 0.033 & 2 & 0.067, 0.068 & $0.24''\times 0.24''$ \\
  ZTF18acbzvpg & SN 2018iby & 94 & 0.06, 0.06 & 0.016, 0.012 & 0.110, 0.090 & 1 & 0.126, 0.109 & $0.44''\times 0.38''$ \\
  ZTF18accnoli & SN 2018hmx & 3490 & 0.06, 0.06 & 0.001, 0.002 & 0.030, 0.060 & 2 & 0.067, 0.085 & $0.23''\times 0.30''$ \\
  ZTF18acqxyiq & SN 2018jak & 33 & 0.06, 0.06 & 0.026, 0.016 & 0.049, 0.062 & 2 & 0.081, 0.088 & $0.28''\times 0.31''$ \\
  ZTF19aadnhdo & SN 2019us & 86 & 0.06, 0.06 & 0.022, 0.028 & 0.030, 0.081 & 3 & 0.071, 0.104 & $0.25''\times 0.37''$ \\
  ZTF19aadttht & SN 2019yz & 11 & 0.06, 0.06 & 0.027, 0.026 & 0.102, 0.032 & 3 & 0.121, 0.073 & $0.42''\times 0.25''$ \\
  ZTF19aakirwj & SN 2019amm & 18 & 0.06, 0.06 & 0.023, 0.022 & 0.208, 0.170 & 2 & 0.218, 0.182 & $0.76''\times 0.64''$ \\
  ZTF19aaknate & SN 2019bao & 23 & 0.06, 0.06 & 0.003, 0.003 & 0.110, 0.090 & 0 & 0.125, 0.108 & $0.44''\times 0.38''$ \\
  ZTF19aakpcuw & SN 2019bgl & 75 & 0.06, 0.06 & 0.013, 0.022 & 0.110, 0.090 & 1 & 0.126, 0.110 & $0.44''\times 0.39''$ \\
  ZTF19aalouag & SN 2019buy & 714 & 0.06, 0.06 & 0.003, 0.005 & 0.110, 0.090 & 1 & 0.125, 0.108 & $0.44''\times 0.38''$ \\
  ZTF19aamgghn & SN 2019bwi & 75 & 0.06, 0.06 & 0.009, 0.013 & 0.102, 0.118 & 2 & 0.119, 0.133 & $0.42''\times 0.46''$ \\
  ZTF19aamsetj & SN 2019cad & 25 & 0.06, 0.06 & 0.015, 0.013 & 0.166, 0.030 & 3 & 0.177, 0.068 & $0.62''\times 0.24''$ \\
  ZTF19aarfkch & SN 2019dwf & 211 & 0.06, 0.06 & 0.010, 0.003 & 0.110, 0.090 & 1 & 0.126, 0.108 & $0.44''\times 0.38''$ \\
  ZTF19aatheus & SN 2019eix & 126 & 0.06, 0.06 & 0.003, 0.008 & 0.062, 0.062 & 17 & 0.086, 0.087 & $0.30''\times 0.30''$ \\
  ZTF19aaxfcpq & SN 2019gwc & 338 & 0.06, 0.06 & 0.005, 0.006 & 0.033, 0.030 & 3 & 0.069, 0.067 & $0.24''\times 0.24''$ \\
  ZTF19abacxod & SN 2019hvg & 86 & 0.06, 0.06 & 0.008, 0.009 & 0.104, 0.062 & 5 & 0.120, 0.087 & $0.42''\times 0.30''$ \\
  ZTF19abafmwj & SN 2019icf & 87 & 0.06, 0.06 & 0.012, 0.013 & 0.055, 0.031 & 7 & 0.082, 0.069 & $0.29''\times 0.24''$ \\
  ZTF19abbthwy & SN 2019igh & 18 & 0.06, 0.06 & 0.015, 0.005 & 0.064, 0.044 & 2 & 0.089, 0.075 & $0.31''\times 0.26''$ \\
  ZTF19abdoior & SN 2019krw & 130 & 0.06, 0.06 & 0.007, 0.009 & 0.066, 0.055 & 9 & 0.090, 0.082 & $0.31''\times 0.29''$ \\
  ZTF19abfsxpw & SN 2019lci & 22 & 0.06, 0.06 & 0.014, 0.005 & 0.040, 0.070 & 15 & 0.074, 0.092 & $0.26''\times 0.32''$ \\
  ZTF19abgfuhh & SN 2019lgc & 69 & 0.06, 0.06 & 0.010, 0.010 & 0.060, 0.030 & 6 & 0.086, 0.068 & $0.30''\times 0.24''$ \\
  ZTF19abgqruu & AT 2019mjo & 29 & 0.06, 0.06 & 0.038, 0.027 & 0.075, 0.075 & 2 & 0.103, 0.100 & $0.36''\times 0.35''$ \\
  ZTF19abqmsnk & SN 2019ofk & 19 & 0.06, 0.06 & 0.011, 0.015 & 0.095, 0.124 & 6 & 0.113, 0.139 & $0.39''\times 0.49''$ \\
  ZTF19abqwtfu & SN 2019odp & 173 & 0.06, 0.06 & 0.004, 0.007 & 0.037, 0.085 & 3 & 0.070, 0.104 & $0.25''\times 0.36''$ \\
  ZTF19abqykei & SN 2019obh & 38 & 0.06, 0.06 & 0.014, 0.021 & 0.057, 0.019 & 5 & 0.084, 0.066 & $0.29''\times 0.23''$ \\
  ZTF19abtsnyy & SN 2019oub & 377 & 0.06, 0.06 & 0.018, 0.013 & 0.048, 0.072 & 11 & 0.079, 0.095 & $0.28''\times 0.33''$ \\
  ZTF19abucwzt & SN 2019oys & 91 & 0.06, 0.06 & 0.011, 0.014 & 0.061, 0.061 & 11 & 0.086, 0.086 & $0.30''\times 0.30''$ \\
  ZTF19abxjrge & SN 2019ply & 37 & 0.06, 0.06 & 0.011, 0.010 & 0.030, 0.079 & 4 & 0.068, 0.100 & $0.24''\times 0.35''$ \\
  ZTF19abxphdh & AT 2019pnm & 47 & 0.06, 0.06 & 0.033, 0.018 & 0.066, 0.080 & 4 & 0.095, 0.101 & $0.33''\times 0.35''$ \\
  ZTF19abxtcio & SN 2019pof & 55 & 0.06, 0.06 & 0.012, 0.011 & 0.054, 0.113 & 7 & 0.082, 0.128 & $0.29''\times 0.45''$ \\
  ZTF19abzwaen & SN 2019qfi & 17 & 0.06, 0.06 & 0.016, 0.012 & 0.031, 0.031 & 5 & 0.069, 0.069 & $0.24''\times 0.24''$ \\
  ZTF19acbvbtz & AT 2019rpt & 24 & 0.06, 0.06 & 0.020, 0.023 & 0.086, 0.061 & 6 & 0.107, 0.089 & $0.37''\times 0.31''$ \\
  ZTF19acfejbj & SN 2019sox & 30 & 0.06, 0.06 & 0.004, 0.021 & 0.071, 0.051 & 11 & 0.093, 0.082 & $0.33''\times 0.29''$ \\
  ZTF19acjtpqd & SN 2019tqb & 5 & 0.06, 0.06 & 0.016, 0.012 & 0.111, 0.054 & 4 & 0.127, 0.081 & $0.45''\times 0.29''$ \\
  ZTF19ackjene & SN 2019tzx & 17 & 0.06, 0.06 & 0.019, 0.018 & 0.110, 0.090 & 0 & 0.127, 0.110 & $0.44''\times 0.38''$ \\
  ZTF19ackjjwf & SN 2019tls & 691 & 0.06, 0.06 & 0.002, 0.008 & 0.024, 0.022 & 7 & 0.065, 0.064 & $0.23''\times 0.23''$ \\
  ZTF19acmelor & SN 2019uff & 15 & 0.06, 0.06 & 0.025, 0.051 & 0.110, 0.090 & 0 & 0.128, 0.119 & $0.45''\times 0.42''$ \\
  ZTF19acxpuql & SN 2019vsi & 714 & 0.06, 0.06 & 0.003, 0.004 & 0.030, 0.030 & 2 & 0.067, 0.067 & $0.24''\times 0.24''$ \\
  ZTF19acyogrm & SN 2019wzj & 27 & 0.06, 0.06 & 0.021, 0.003 & 0.040, 0.082 & 8 & 0.075, 0.101 & $0.26''\times 0.35''$ \\
  ZTF19adbrjic & SN 2019xuk & 333 & 0.06, 0.06 & 0.001, 0.000 & 0.030, 0.030 & 2 & 0.067, 0.067 & $0.23''\times 0.23''$ \\
  ZTF20aaaxghv & SN 2019yfv & 8 & 0.06, 0.06 & 0.020, 0.032 & 0.066, 0.069 & 3 & 0.092, 0.097 & $0.32''\times 0.34''$ \\
  ZTF20aaelulu & SN 2020oi & 111 & 0.06, 0.06 & 0.005, 0.007 & 0.037, 0.007 & 8 & 0.071, 0.061 & $0.25''\times 0.21''$ \\
  ZTF20aaerxne & SN 2019yxp & 82 & 0.06, 0.06 & 0.009, 0.013 & 0.054, 0.030 & 3 & 0.081, 0.068 & $0.28''\times 0.24''$ \\
  ZTF20aafmdzj & SN 2020zg & 26 & 0.06, 0.06 & 0.038, 0.008 & 0.110, 0.090 & 1 & 0.131, 0.108 & $0.46''\times 0.38''$ \\
  ZTF20aahbamv & SN 2020amv & 370 & 0.06, 0.06 & 0.002, 0.004 & 0.057, 0.119 & 3 & 0.083, 0.133 & $0.29''\times 0.47''$ \\
  ZTF20aahdutc & SN 2020akf & 20 & 0.06, 0.06 & 0.002, 0.015 & 0.030, 0.030 & 2 & 0.067, 0.069 & $0.23''\times 0.24''$ \\
  ZTF20aahfqpm & SN 2020ano & 26 & 0.06, 0.06 & 0.059, 0.001 & 0.030, 0.030 & 4 & 0.089, 0.067 & $0.31''\times 0.23''$ \\
  ZTF20aaiqiti & SN 2020ayz & 256 & 0.06, 0.06 & 0.008, 0.012 & 0.063, 0.032 & 3 & 0.087, 0.069 & $0.31''\times 0.24''$ \\
  ZTF20aalcyih & SN 2020bpf & 38 & 0.06, 0.06 & 0.011, 0.017 & 0.054, 0.046 & 20 & 0.081, 0.077 & $0.28''\times 0.27''$ \\
  ZTF20aammtwx & SN 2020cgu & 66 & 0.06, 0.06 & 0.009, 0.005 & 0.016, 0.046 & 14 & 0.063, 0.076 & $0.22''\times 0.27''$ \\
  ZTF20aanvmdt & SN 2020cpg & 16 & 0.06, 0.06 & 0.012, 0.007 & 0.033, 0.030 & 3 & 0.070, 0.067 & $0.24''\times 0.24''$ \\
  ZTF20aavgmli & SN 2020htt & 20 & 0.06, 0.06 & 0.022, 0.010 & 0.143, 0.060 & 3 & 0.157, 0.085 & $0.55''\times 0.30''$ \\
  ZTF20abaszgh & SN 2020ksa & 57 & 0.06, 0.06 & 0.012, 0.009 & 0.315, 0.117 & 2 & 0.321, 0.132 & $1.12''\times 0.46''$ \\
  ZTF20abbplei & SN 2020lao & 83 & 0.06, 0.06 & 0.013, 0.008 & 0.048, 0.102 & 8 & 0.078, 0.119 & $0.27''\times 0.41''$ \\
  ZTF20abdajuq & AT 2020lma & 43 & 0.06, 0.06 & 0.017, 0.014 & 0.032, 0.030 & 2 & 0.070, 0.069 & $0.25''\times 0.24''$ \\
  ZTF20abevbxv & SN 2020mmz & 46 & 0.06, 0.06 & 0.006, 0.013 & 0.030, 0.212 & 2 & 0.067, 0.221 & $0.24''\times 0.77''$ \\
  ZTF20abgbuly & SN 2020jfv & 309 & 0.06, 0.06 & 0.006, 0.006 & 0.086, 0.096 & 4 & 0.105, 0.113 & $0.37''\times 0.40''$ \\
  ZTF20abgbxvb & SN 2020nac & 30 & 0.06, 0.06 & 0.005, 0.017 & 0.089, 0.072 & 21 & 0.107, 0.095 & $0.38''\times 0.33''$ \\
  ZTF20abhlncz & SN 2020nke & 64 & 0.06, 0.06 & 0.012, 0.009 & 0.079, 0.081 & 6 & 0.100, 0.101 & $0.35''\times 0.36''$ \\
  ZTF20abkiarz & SN 2020oce & 20 & 0.06, 0.06 & 0.005, 0.010 & 0.026, 0.119 & 9 & 0.066, 0.134 & $0.23''\times 0.47''$ \\
  ZTF20abodalh & SN 2020qbu & 13 & 0.06, 0.06 & 0.042, 0.086 & 0.072, 0.077 & 4 & 0.103, 0.130 & $0.36''\times 0.46''$ \\
  ZTF20abwzqzo & SN 2020sbw & 46 & 0.06, 0.06 & 0.010, 0.016 & 0.067, 0.030 & 2 & 0.091, 0.069 & $0.32''\times 0.24''$ \\
  ZTF20abxpoxd & SN 2020sgf & 192 & 0.06, 0.06 & 0.003, 0.006 & 0.032, 0.029 & 14 & 0.068, 0.067 & $0.24''\times 0.23''$ \\
  ZTF20abxyaaf & SN 2020rur & 31 & 0.06, 0.06 & 0.002, 0.006 & 0.042, 0.049 & 9 & 0.073, 0.077 & $0.26''\times 0.27''$ \\
  ZTF20abywsut & SN 2020sya & 60 & 0.06, 0.06 & 0.008, 0.008 & 0.079, 0.064 & 3 & 0.100, 0.088 & $0.35''\times 0.31''$ \\
  ZTF20acjpvyd & SN 2020tjd & 58 & 0.06, 0.06 & 0.007, 0.012 & 0.040, 0.064 & 9 & 0.072, 0.089 & $0.25''\times 0.31''$ \\
  ZTF20aclvvws & SN 2020ybq & 21 & 0.06, 0.06 & 0.015, 0.021 & 0.041, 0.078 & 14 & 0.074, 0.101 & $0.26''\times 0.35''$ \\
  ZTF20acpjqxp & SN 2020zgl & 64 & 0.06, 0.06 & 0.008, 0.009 & 0.030, 0.030 & 2 & 0.068, 0.068 & $0.24''\times 0.24''$ \\
  ZTF20actqnhg & SN 2020aaxf & 68 & 0.06, 0.06 & 0.010, 0.022 & 0.060, 0.100 & 7 & 0.085, 0.119 & $0.30''\times 0.42''$ \\
  ZTF20acwobku & SN 2020acct & 102 & 0.06, 0.06 & 0.008, 0.008 & 0.110, 0.090 & 1 & 0.126, 0.108 & $0.44''\times 0.38''$ \\
  ZTF20acwofhd & SN 2020acfp & 73 & 0.06, 0.06 & 0.021, 0.017 & 0.153, 0.068 & 2 & 0.165, 0.093 & $0.58''\times 0.32''$ \\
  ZTF20acwqqjs & SN 2020acat & 39 & 0.06, 0.06 & 0.007, 0.007 & 0.030, 0.030 & 2 & 0.067, 0.067 & $0.24''\times 0.24''$ \\
  ZTF20adadrhw & SN 2020adow & 185 & 0.06, 0.06 & 0.003, 0.006 & 0.138, 0.125 & 2 & 0.150, 0.138 & $0.53''\times 0.48''$ \\
  ZTF21aaabrpu & SN 2021rf & 83 & 0.06, 0.06 & 0.007, 0.014 & 0.066, 0.067 & 2 & 0.089, 0.091 & $0.31''\times 0.32''$ \\
  ZTF21aaaosoq & SN 2021bm & 43 & 0.06, 0.06 & 0.015, 0.008 & 0.112, 0.054 & 4 & 0.128, 0.081 & $0.45''\times 0.28''$ \\
  ZTF21aaaubig & SN 2021do & 32 & 0.06, 0.06 & 0.007, 0.006 & 0.110, 0.090 & 1 & 0.126, 0.108 & $0.44''\times 0.38''$ \\
  ZTF21aacufip & SN 2021vz & 55 & 0.06, 0.06 & 0.038, 0.015 & 0.056, 0.114 & 4 & 0.091, 0.129 & $0.32''\times 0.45''$ \\
  ZTF21aaoiaar & SN 2021esi & 43 & 0.06, 0.06 & 0.015, 0.018 & 0.056, 0.069 & 17 & 0.084, 0.093 & $0.29''\times 0.33''$ \\
  ZTF21aapfnoc & AT 2021fyq & 17 & 0.06, 0.06 & 0.052, 0.040 & 0.110, 0.090 & 0 & 0.136, 0.115 & $0.47''\times 0.40''$ \\
  ZTF21aaqvsvw & SN 2021heh & 36 & 0.06, 0.06 & 0.006, 0.016 & 0.049, 0.041 & 7 & 0.078, 0.075 & $0.27''\times 0.26''$ \\
  ZTF21aarnjyd & SN 2021hrj & 217 & 0.06, 0.06 & 0.006, 0.006 & 0.110, 0.090 & 1 & 0.125, 0.108 & $0.44''\times 0.38''$ \\
\enddata
\tablecomments{All uncertainties are in arcseconds. $\sigma_{\rm ZTF,sys}$: ZTF systematic astrometric uncertainty per axis \citep{Masci2019}. $\sigma_{\rm ZTF,stat}$: statistical scatter of ZTF alert positions around nightly weighted centroids. $\sigma_{\rm HST}$: HST WCS uncertainty per axis, measured empirically from Gaia~DR3 using the canonical visit DRC plus the two individual FLC subexposures when at least two good Gaia stars are recovered; otherwise the header WCS catalog is used. $\sigma_{\rm tot}$: total per-axis uncertainty (all terms added in quadrature). $3.5\sigma$ search: semi-axes of the $3.5\sigma$ error ellipse used for source detection.}
\end{deluxetable*}

\bibliography{references_main}

\begin{figure*}
    \centering
    \includegraphics[width=1.0\linewidth,page=1]{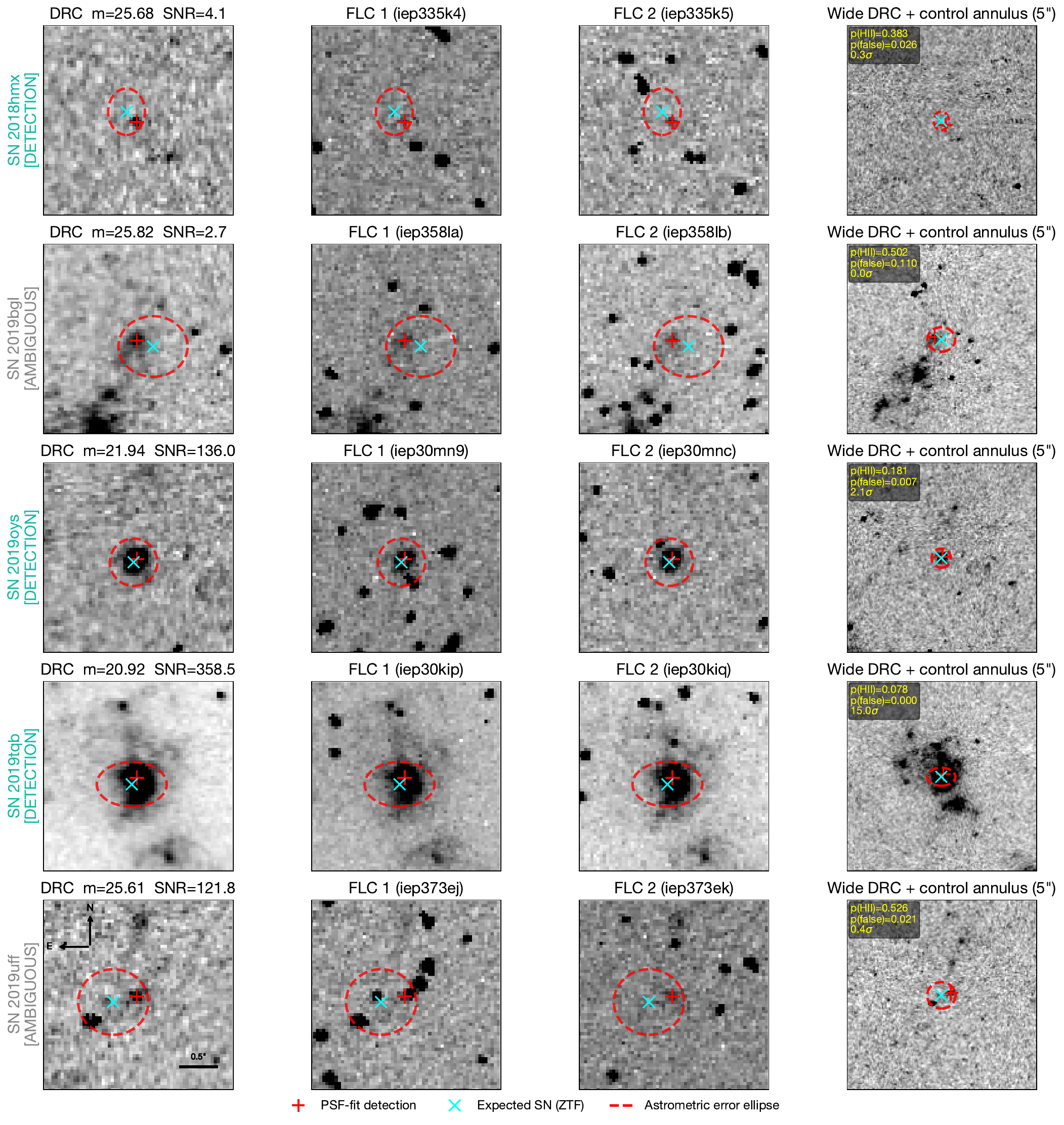}
    \caption{Aligned HST diagnostics for all secure and ambiguous UV candidate counterparts. For each target the first three panels show the DRC image and the two individual FLC exposures on a common astrometric grid centered on the adopted primary-candidate position; the fourth panel shows the wider DRC environment and the control annulus used for the local source-density comparison. The red plus marks the adopted primary-candidate position and the cyan symbol marks the expected SN position from the astrometric solution. The wide-field annotations list the empirical host-confusion diagnostics $p(\mathrm{HII})$, the local false-association probability $p(\mathrm{false})$, and $\sigma_{\rm control}$. Here $p(\mathrm{false})$ is the fraction of random same-size ellipses in the local control annulus that contain any source at least as bright as the adopted candidate, regardless of where it lands within the ellipse. At present only $p(\mathrm{HII})$ enters the formal downgrade-to-ambiguous check; $p(\mathrm{false})$ is shown as a diagnostic-only metric.}
    \label{fig:diag_candidates}
\end{figure*}

\begin{figure*}
    \centering
    \includegraphics[width=1.0\linewidth,page=2]{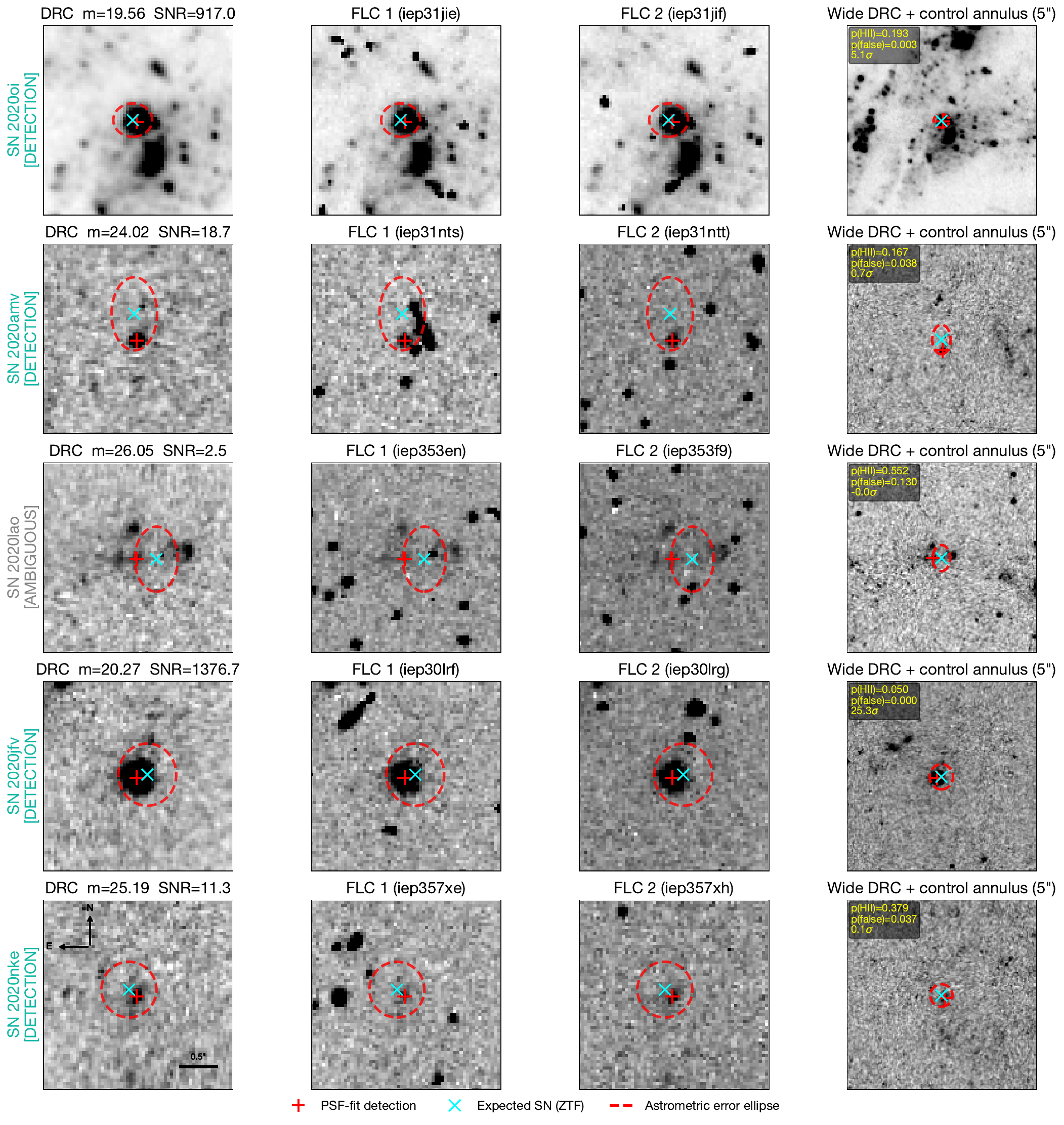}
    \caption{Continued from Figure~\ref{fig:diag_candidates}.}
\end{figure*}

\begin{figure*}
    \centering
    \includegraphics[width=1.0\linewidth,page=3]{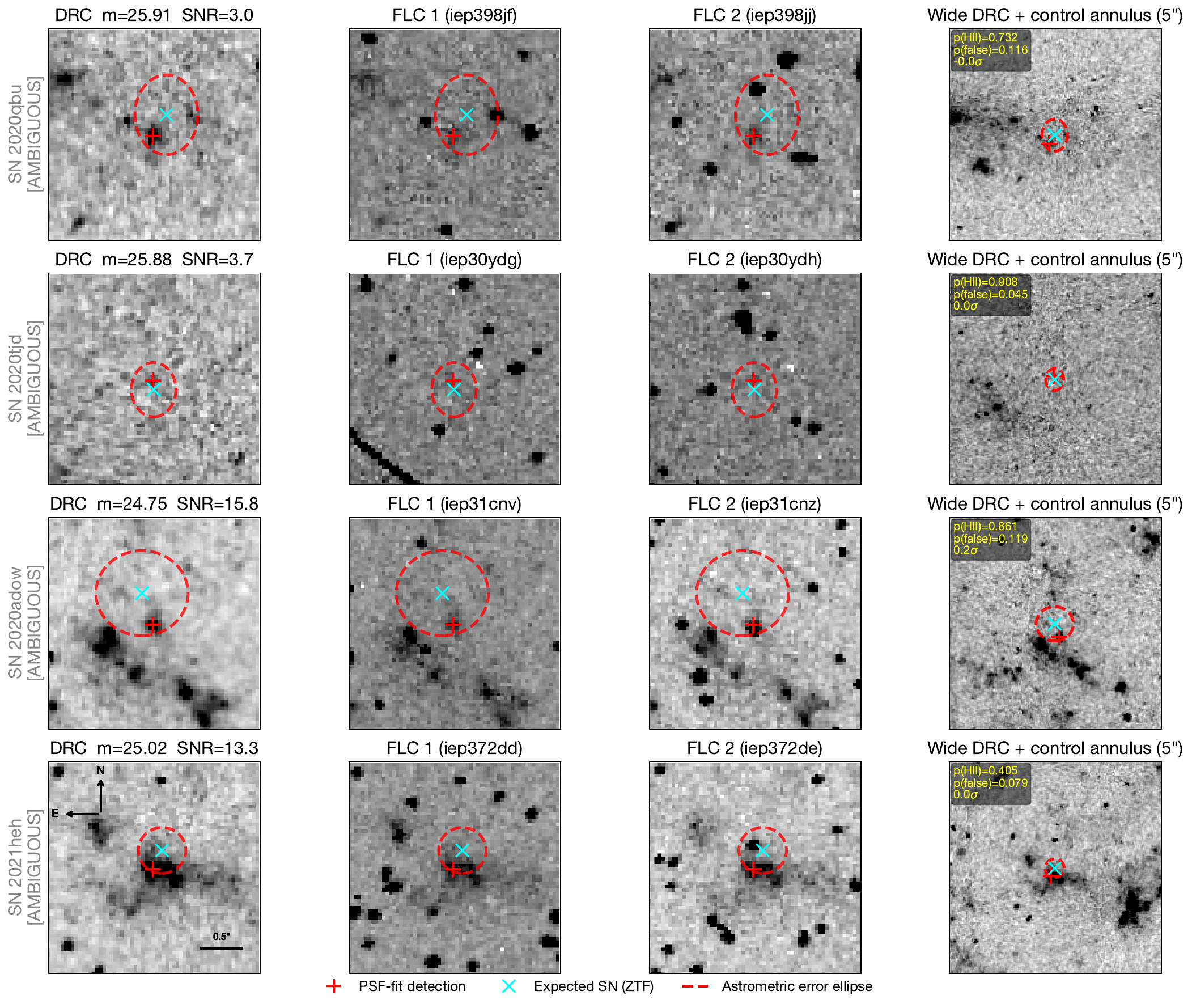}
    \caption{Continued from Figure~\ref{fig:diag_candidates}.}
\end{figure*}

\begin{figure*}
    \centering
    \includegraphics[width=1.0\linewidth]{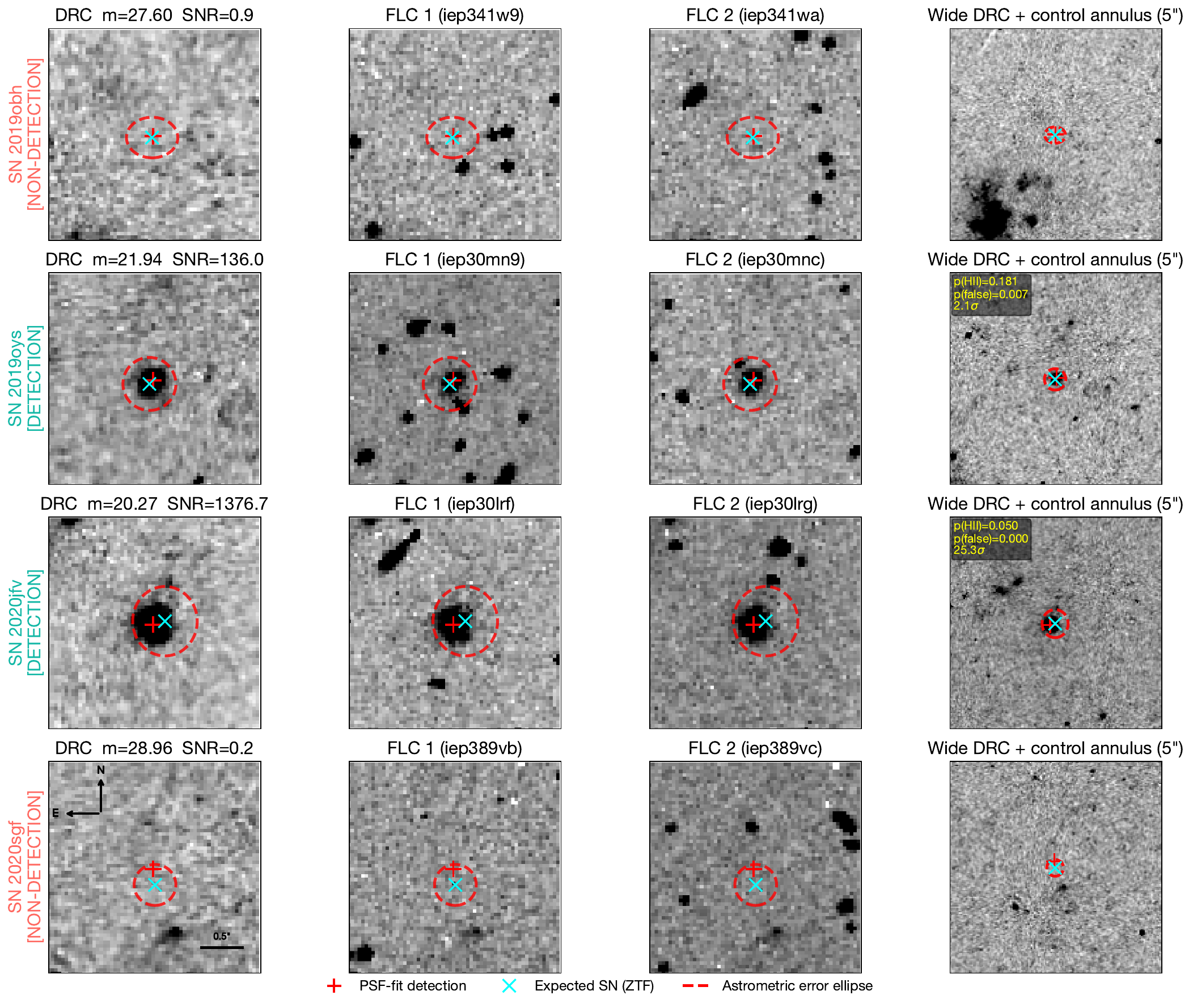}
    \caption{Aligned HST diagnostics for the four robust WISE-selected interaction candidates: SN~2019obh, SN~2019oys, SN~2020jfv, and SN~2020sgf. The panel layout matches Figure~\ref{fig:diag_candidates}. This figure is useful because the WISE-selected subset includes both secure HST UV counterparts and HST non-detections.}
    \label{fig:diag_wise}
\end{figure*}

\begin{figure*}
    \centering
    \includegraphics[width=1.0\linewidth,page=1]{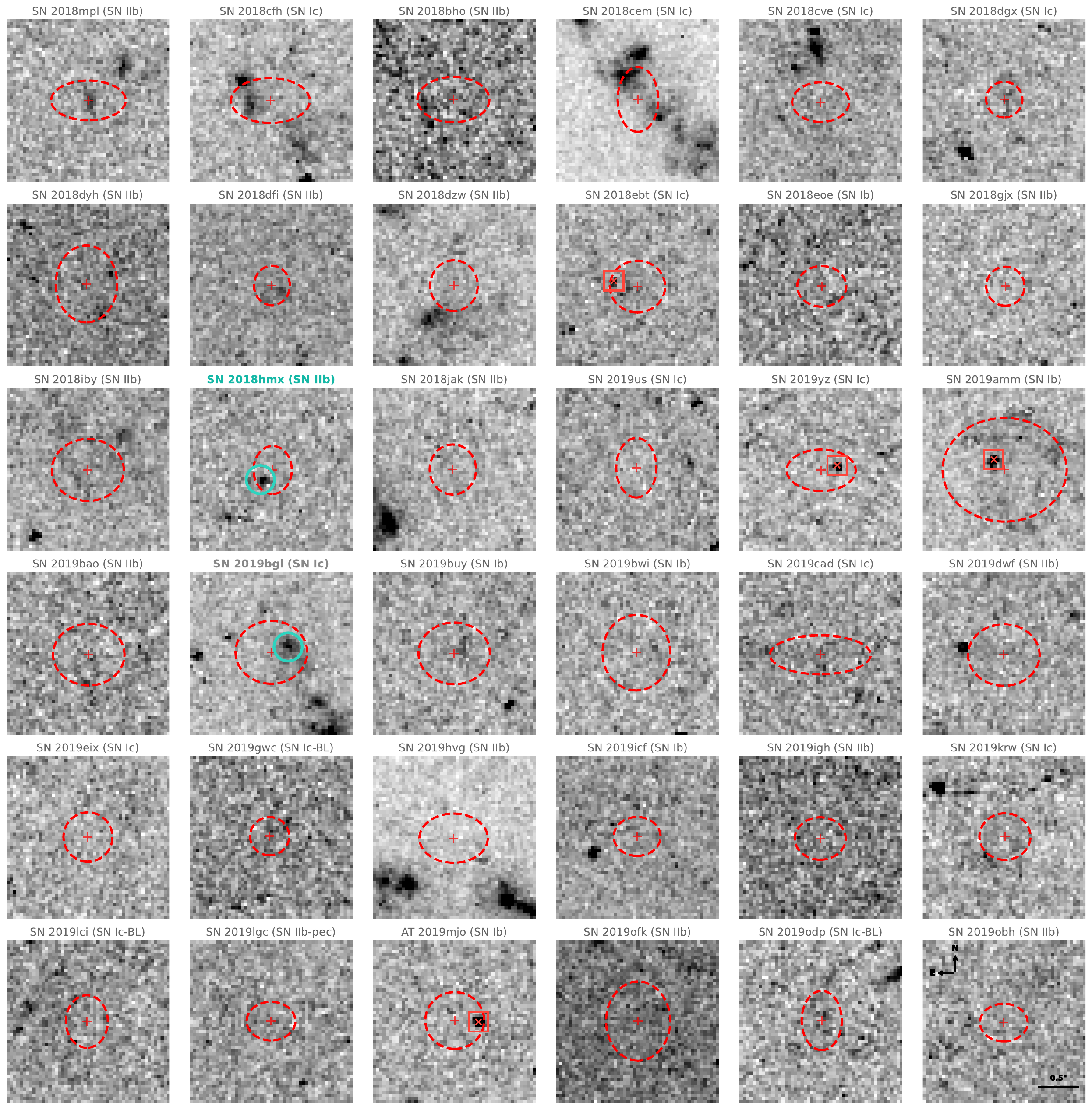}
    \caption{HST WFC3/UVIS cutouts for the full survey sample. Each panel shows a $1\farcs0$ radius region centered on the expected SN position. Red dashed ellipses mark the $3.5\sigma$ astrometric uncertainty region (Table~\ref{tab:astrometry}). A thick solid cyan circle marks the adopted primary counterpart when one is selected, while thinner dashed cyan circles mark additional FLC-confirmed in-ellipse sources. Red squares mark sources rejected by the FLC comparison as cosmic rays. Object names are annotated with their spectroscopic type. Confirmed detections ($\ast$) and ambiguous candidates ($\sim$) are color-coded.}
    \label{fig:all_cutouts}
\end{figure*}

\begin{figure*}
    \centering
    \includegraphics[width=1.0\linewidth,page=2]{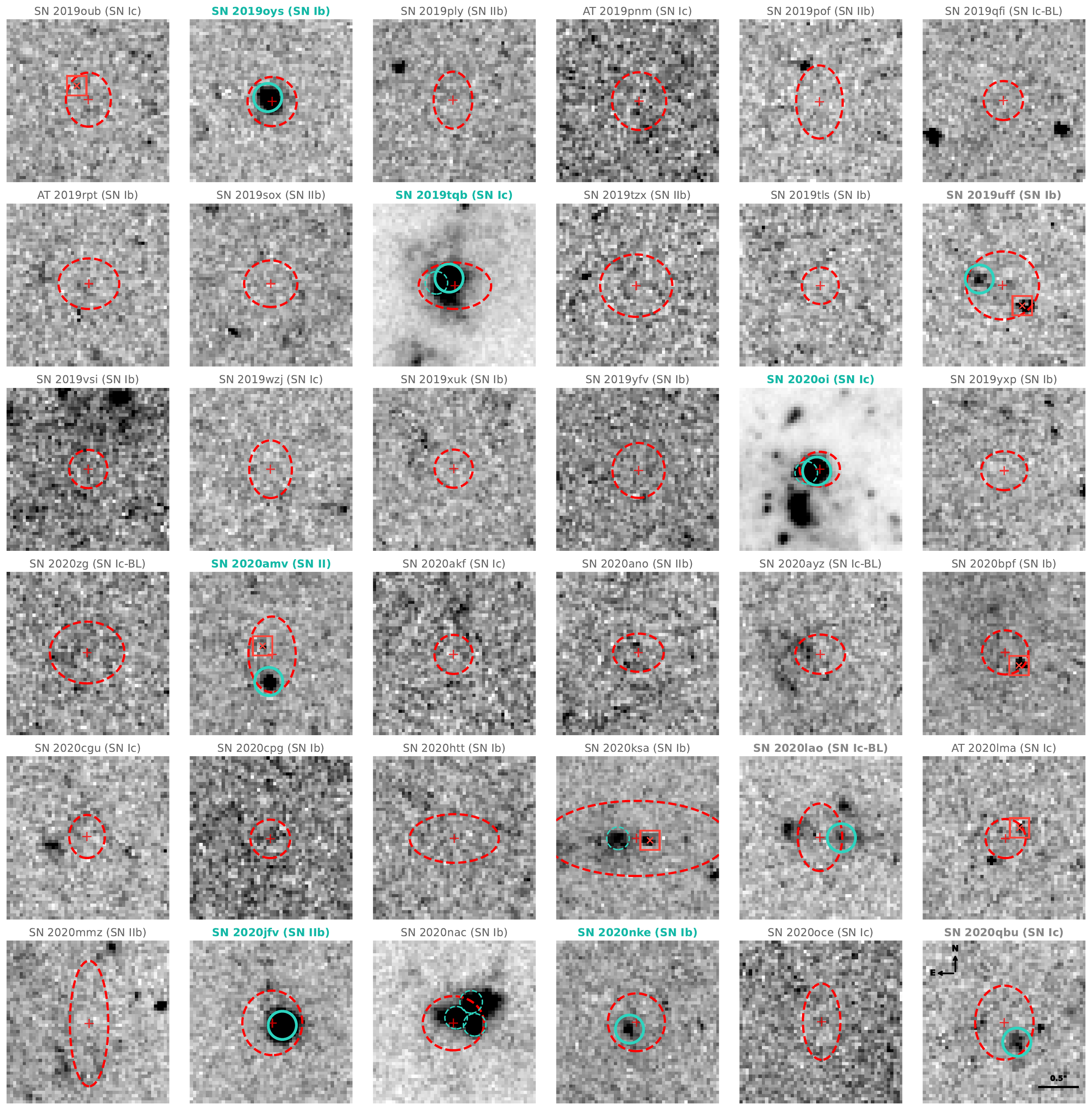}
    \caption{Continued from Figure~\ref{fig:all_cutouts}.}
\end{figure*}

\begin{figure*}
    \centering
    \includegraphics[width=1.0\linewidth,page=3]{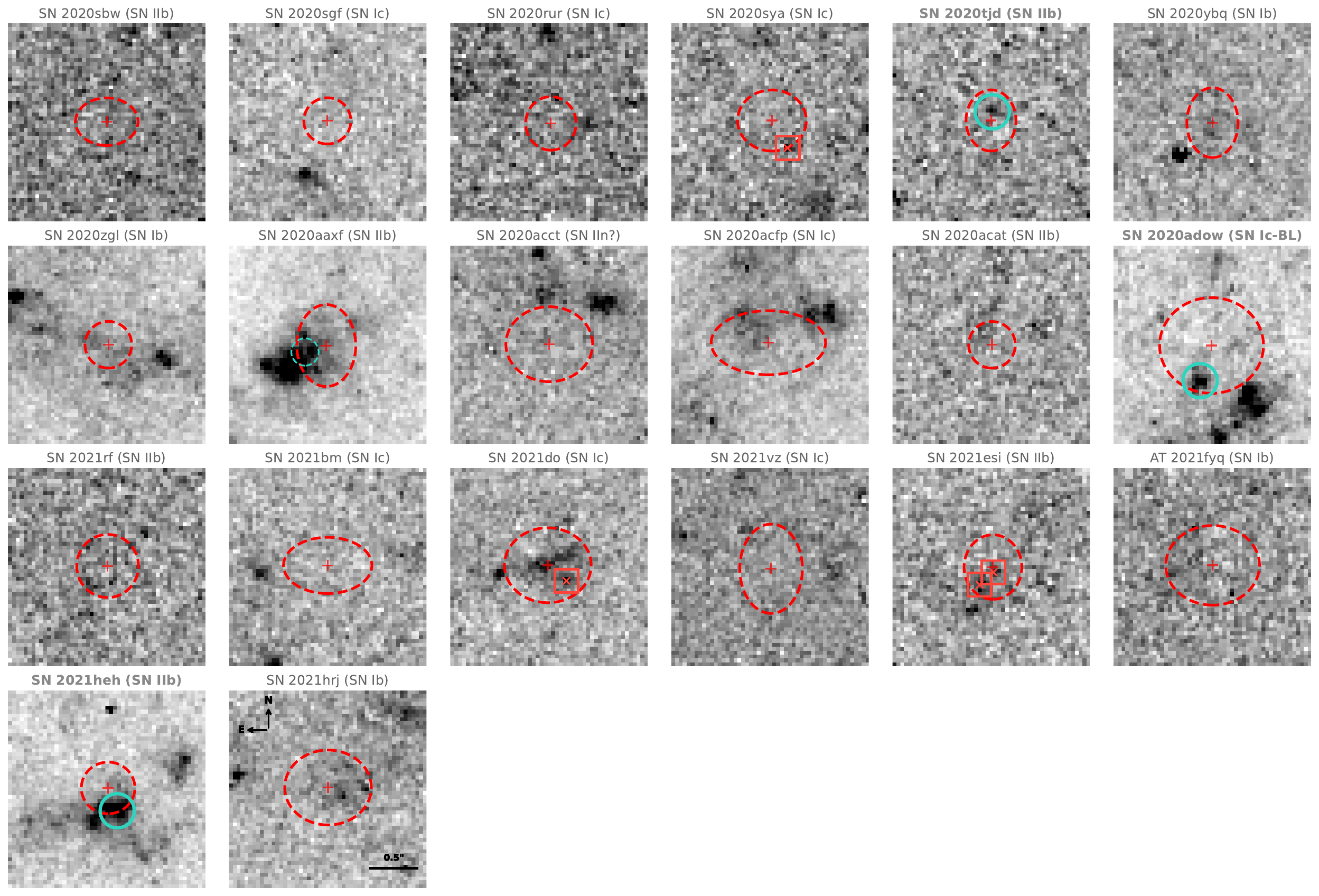}
    \caption{Continued from Figure~\ref{fig:all_cutouts}.}
\end{figure*}


\section{Survey Detection Efficiency}\label{sec:efficiency}

\begin{figure}
    \centering
    \includegraphics[width=\columnwidth]{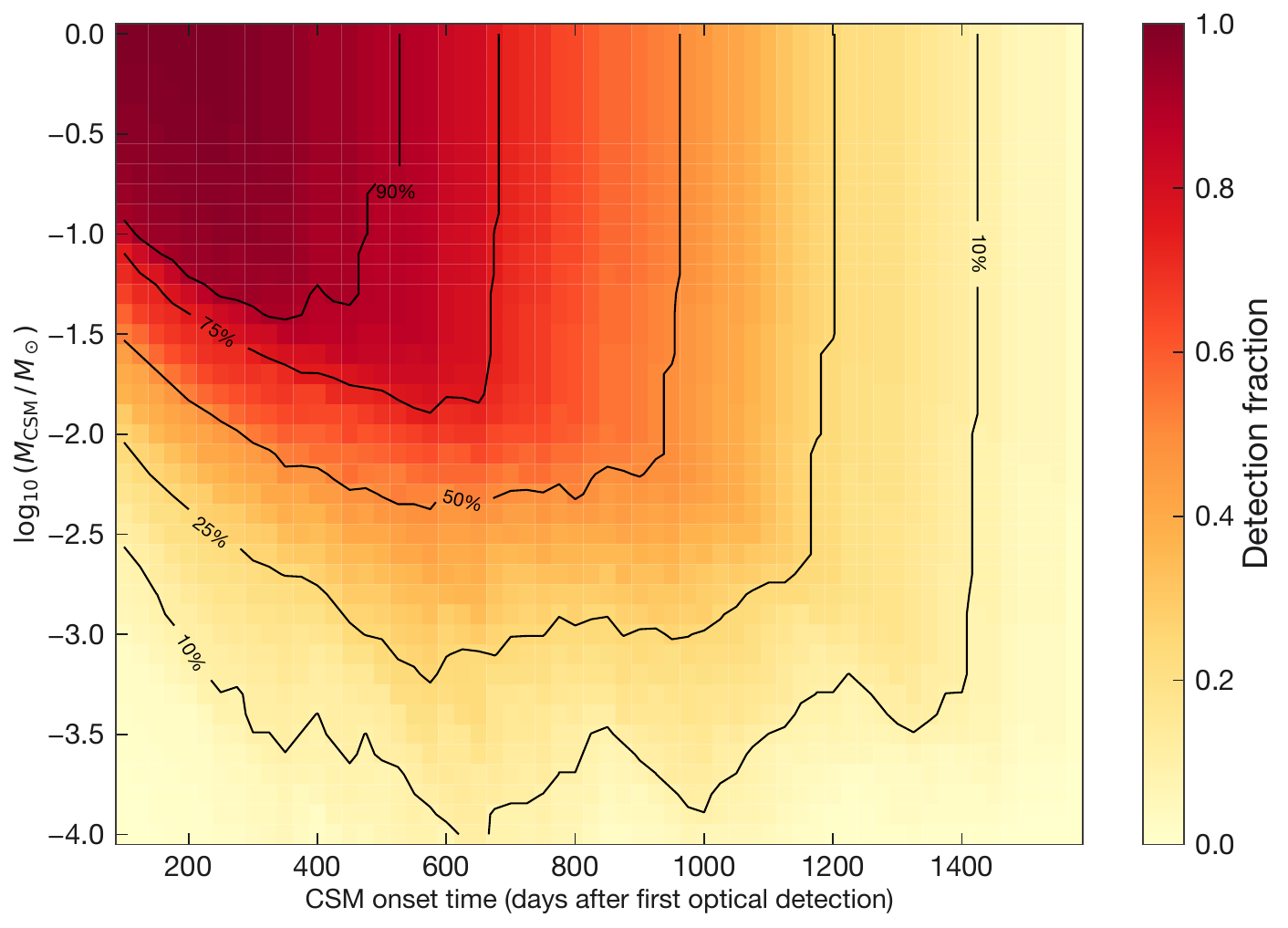}
    \caption{Detection efficiency of the survey as a function of CSM shell mass and interaction onset time, computed using the CSM physics model (Sect.~\ref{sec:csmmodel}) with the posterior median shell thickness ($f_\mathrm{thick} = \fthickmed$). The color scale shows the fraction of the \Ntotal\ survey targets for which a CSM episode with the given $M_\mathrm{CSM}$ and $t_\mathrm{onset}$ would produce UV emission brighter than the adopted $70\%$-coverage injection/recovery limit at the target's observation epoch. Black contours mark 10\%, 25\%, 50\%, 75\%, and 90\% detection fractions.}
    \label{fig:efficiency}
\end{figure}

Figure~\ref{fig:efficiency} shows the detection efficiency of the survey as a function of CSM shell mass and interaction onset time, computed using the CSM physics model (Sect.~\ref{sec:csmmodel}) with the posterior median shell thickness ($f_\mathrm{thick} = \fthickmed$). For each point on the $(M_\mathrm{CSM},\, t_\mathrm{onset})$ grid, the model predicts the UV absolute magnitude at each target's observation epoch, and the detection fraction is the fraction of targets for which that predicted magnitude falls above the adopted $70\%$-coverage injection/recovery limit. The survey is $>90\%$ complete for $M_\mathrm{CSM} \gtrsim 0.03~M_\odot$ at onset times $\lesssim 400$~days, and reaches $\sim$50\% completeness near the posterior median shell mass ($M_\mathrm{CSM} \approx \MCSMsolmed~M_\odot$). At late onset times ($\gtrsim 1000$~days), only the most massive shells remain detectable because most targets were observed before interaction would have begun. This visualization complements but does not replace the MCMC analysis, which marginalizes over onset time and uses the observed magnitudes of detections rather than binary detection/non-detection.

\section{Simulated Survey Realization}\label{sec:simulated}

\begin{figure*}
    \centering
    \includegraphics[width=1.0\linewidth]{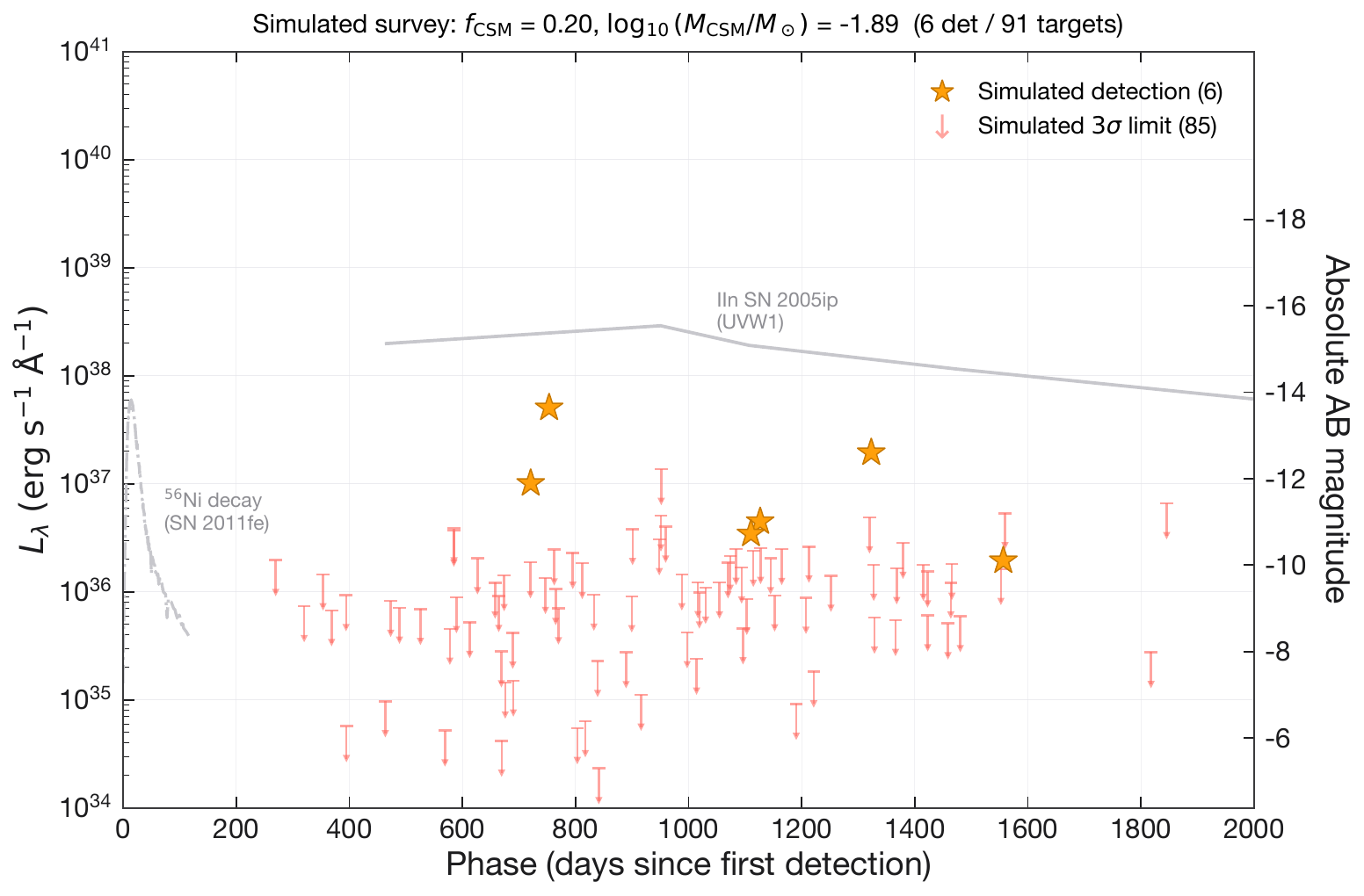}
    \caption{Simulated survey realization using the median posterior CSM interaction model ($f_\mathrm{CSM} = \fCSMmed$, $\log_{10}(M_\mathrm{CSM}/M_\odot) = \logMCSMmed$, $\log_{10} f_\mathrm{thick} = \logfthickmed$). Each of the \Ntotal\ real targets is assigned a CSM episode with probability $f_\mathrm{CSM}$ at a random onset time, and the resulting UV magnitude is compared to the adopted $70\%$-coverage injection/recovery limit. Compare to Figure~\ref{fig:results}.}
    \label{fig:simulated}
\end{figure*}

As a consistency check, Figure~\ref{fig:simulated} shows a single Monte Carlo realization of the survey using the median posterior CSM physics model (Sect.~\ref{sec:mcmc}). We assign each of the \Ntotal\ targets a CSM episode with probability $f_\mathrm{CSM}$, draw a random interaction onset time uniformly from $[0,\, 2000]$~days, compute the resulting UV luminosity from the physics model with the median posterior shell mass and 0.5~mag intrinsic scatter, and compare to each target's adopted $70\%$-coverage injection/recovery limit. The phase, redshift, and limiting magnitude of every target are taken from the real survey. The simulated detections span a comparable range of luminosities and phases as Figure~\ref{fig:results}, confirming that the physics model reproduces the global characteristics of the observed sample.

\end{document}